\PassOptionsToPackage{table,dvipsnames}{xcolor}

\documentclass[10pt,twocolumn,letterpaper]{article}

\usepackage{cvpr}              %
\usepackage[accsupp]{axessibility}  %

\definecolor{cvprblue}{rgb}{0.21,0.49,0.74}
\usepackage[pagebackref,breaklinks,colorlinks,allcolors=cvprblue]{hyperref}

\definecolor{yellow}{rgb}{1,1, 0.8}
\definecolor{orange}{rgb}{1, 0.85, 0.7}

\makeatletter
\renewcommand{\paragraph}{%
  \@startsection{paragraph}{4}%
  {\z@}{0.2ex \@plus 0.3ex \@minus .2ex}{-1em}%
  {\normalfont\normalsize\bfseries}%
}
\makeatother
\def\paperID{1684} %
\def\confName{CVPR}
\def\confYear{2025}

\newcommand{\sysName}{SketchVideo}
\title{\sysName: Sketch-based Video Generation and Editing }

\author{
Feng-Lin Liu\textsuperscript{1,2}~~~
Hongbo Fu\textsuperscript{3}~~~
Xintao Wang\textsuperscript{4}~~~
Weicai Ye\textsuperscript{4}~~~
Pengfei Wan\textsuperscript{4}~~~
Di Zhang\textsuperscript{4}~~~
Lin Gao\textsuperscript{1,2*}
\vspace{2mm}
\\
\parbox{\textwidth}{\centering \small
{\textsuperscript{1}Beijing Key Laboratory of Mobile Computing and Pervasive Device, Institute of Computing Technology, Chinese Academy of Sciences}\\
{\textsuperscript{2}University of Chinese Academy of Sciences}\quad
{\textsuperscript{3}Hong Kong University of Science and Technology}\quad
{\textsuperscript{4}Kuaishou Technology}\quad
\vspace{2mm}
}
}

\begin{document}

\twocolumn[{
    \renewcommand\twocolumn[1][]{#1}%
    \maketitle

    \vspace{-12mm}
    
    \begin{center}
        \centering
        \includegraphics[width=0.94\textwidth]{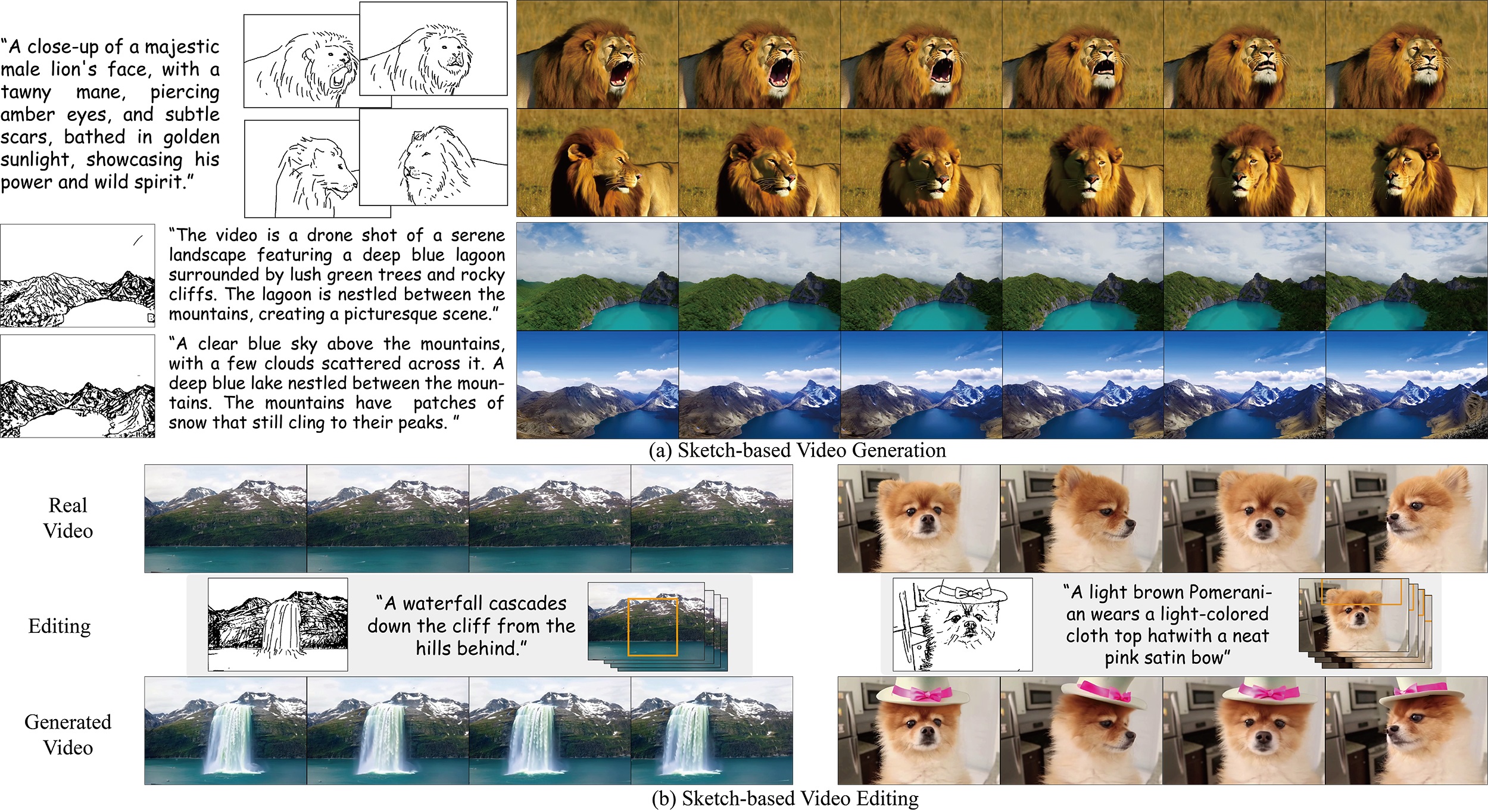}
        \vspace{-3pt}
        \captionof{figure}{
        Our method enables high-quality video generation (a) and editing (b) based on sketch and text inputs. 
        (a) Top: With the same text prompt, different keyframe sketches lead to results with similar semantics but diverse sketch-faithful geometry.
        (a) Bottom: With the same sketches, varied text prompts yield diverse appearances.
        (b) Users can also edit real videos by drawing on keyframe sketches, with edits automatically propagated even when edited objects in original videos have translation and rotation.
        }
        \label{teaser}
    \end{center}
}]

\let\thefootnote\relax\footnotetext{*Corresponding Author: Lin Gao (gaolin@ict.ac.cn)}

\begin{abstract}
Video generation and editing conditioned on text prompts or images have undergone significant advancements.
However, challenges remain in accurately controlling global layout and geometry details solely by texts, and supporting motion control and local modification through images. 
In this paper, we aim to achieve sketch-based spatial and motion control for video generation and support fine-grained editing of real or synthetic videos. 
Based on the DiT video generation model, we propose a memory-efficient control structure with sketch control blocks that predict residual features of skipped DiT blocks. 
Sketches are drawn on one or two keyframes (at arbitrary time points) for easy interaction.
To propagate such temporally sparse sketch conditions across all frames, we propose an inter-frame attention mechanism to analyze the relationship between the keyframes and each video frame.
For sketch-based video editing, we design an additional video insertion module that maintains consistency between the newly edited content and the original video's spatial feature and dynamic motion.
During inference, we use latent fusion for the accurate preservation of unedited regions. 
Extensive experiments demonstrate that our {\sysName} achieves superior performance in controllable video generation and editing. 
Homepage and code: \href{http://geometrylearning.com/SketchVideo/}{http://geometrylearning.com/SketchVideo/}

\end{abstract}
    
\section{Introduction}
\label{sec:intro}

Diffusion-based text-to-image~\cite{StableDiffusion, PixArt, DeepFloyd} and text-to-video ~\cite{sora, opensora, Open-Sora-Plan, cogvideo, cogvideox} models advance significantly due to improvements in datasets~\cite{LAION_5B, COYO_700M, OpenVid_1M, Sketch_jig_dataset} and denoising network architectures~\cite{StableDiffusion, PixArt}.
While text prompts effectively describe high-level semantics, they lack control of scene layouts and geometric details.
To address this, existing video generation methods~\cite{SVD, LVDM, cogvideox} utilize images as additional conditions but raise the questions of how to generate input images and achieve detailed editing.
Sketching serves as a user-friendly interaction tool to capture spatial content and shape details accurately.
One or two sketches are already sufficient to convey desired scene structures and motion information for short video clips (around 6 seconds), which are the target of our and most existing video generation methods.
However, using such sparse keyframe sketches presents several challenges, including reasonably completing the missing frames, %
improving memory efficiency, and addressing the limited size of video datasets.

A na\"ive solution is to translate the input keyframe sketches into images and then utilize interpolation methods~\cite{AMT, Seine, ToonCrafter} for video generation.
However, it is nontrivial to ensure consistency during keyframe sketch-to-image generation, %
which significantly affects the video quality.
This approach {also} struggles to generate extrapolation frames when applying conditions in intermediate frames rather than beginning and ending time points.
Another possible approach is to utilize white placeholders to fill missing condition frames and directly apply ControlNet~\cite{ControlNet} into video models, similar to SparseCtrl~\cite{SparseCtrl}.
However, this requires the same network to process both sketches and white placeholders simultaneously, while the pretrained blocks handle tasks far from this sparse propagation.
Additionally, for DiT-based video frameworks~\cite{cogvideo, opensora}, the %
traditional strategy~\cite{PIXART_delta} that copies half of the pretrained model as a condition network easily causes the out-of-memory issue.

To address these issues, we propose a novel sketch condition network specifically designed for the DiT-based video generation architecture (CogVideoX~\cite{cogvideox} in our work).
Following ControlNet~\cite{ControlNet, PIXART_delta}, we employ a trainable copy of CogVideoX’s DiT block to process only the sketch inputs and generate control features.
No white placeholder is processed to align with the pretrained weights and reduce learning complexity.
To propagate these keyframe features, we design an inter-frame attention mechanism that captures the relationship between the control keyframes and all video frames.
Our approach computes query and key features from noisy latent while extracting value features from sketch conditions. This design leverages frame-to-frame similarity for control propagation.
The above components consist of a single sketch control block.
Instead of copying a half number of the pretrained blocks to construct the control network~\cite{PIXART_delta}, we use only 5 sketch control blocks out of the 30 DiT blocks available in CogVideoX-2b.
We design a novel uniformly distributed skip structure to add the control signals to different levels of features in discrete blocks (0, 6, 12, 18, and 24), achieving effective spatial control while improving memory efficiency.
During training, an external image dataset is incorporated to solve the challenge of limited video data.

Beyond generation, interactive editing of real or synthetic videos further enhances creative flexibility.
Existing methods~\cite{Rerender_A_Video, I2VEdit, AnyV2V, ReVideo} achieve interesting text-based editing or effectively propagate single image editing into videos.
Despite their effectiveness in appearance modification, they struggle with shape manipulation and object insertion, as they preserve the original temporal motion. Such information is missing for newly introduced content.
Moreover, precisely identifying and preserving unedited regions for localized editing remains a challenge.

We propose a sketch-based editing method for detailed local modification. 
Rather than editing a single image and propagating changes, we directly construct an editing network based on our sketch control network. 
To analyze the relationship between edited regions and the original video, we incorporate a video insertion module that takes the original video with masked edited regions as inputs. The modified sketch control blocks generate residual features that capture temporally and spatially coherent contents in the edited areas. 
{To accurately preserve unedited regions and achieve seamless fusion, the newly edited regions are blended with the original video in the latent space.
}

Extensive experiments demonstrate that our method outperforms existing approaches in video generation and editing. 
Our contributions are summarized as follows:
1) We propose \sysName, a novel sketch-based video generation and editing framework that enables detailed geometry control and manipulation using keyframe sketches, as shown in Fig. \ref{teaser}.
2) We design a sketch condition network that predicts skipped control features for the DiT framework, with an inter-frame attention mechanism to propagate one or two sketch conditions across the video.
3) We propose a video insertion module that analyzes the relationship between drawn sketches and original videos, utilizing a latent fusion strategy to preserve unedited regions accurately.

\section{Related Work}

\textbf{Sketch-based Image Generation.}
GAN-based methods~\cite{GAN} have achieved great success in category-restricted sketch-to-image translation ~\cite{pix2pixHD, CycleGAN, DeepFaceDrawing, DeepFaceEditing, pSp, Sketch_jig}. 
Recently, diffusion-based text-to-image models~\cite{StableDiffusion, DeepFloyd} handle general categories with conditional models like ControlNet~\cite{ControlNet}, T2I-Adapter~\cite{T2I-Adapter} and further advances~\cite{Unicontrol, One_Step_Translation, ControlNeXt, DiffMat, control_diffusion_survey}.
Beyond U-Net, the DiT backbone \cite{PixArt} enables image generation, with PIXART-{\(\delta\)}~\cite{PIXART_delta} utilizing the first half of the pretrained model's blocks as a trainable network to predict corresponding control residual features.
{Video generation, however, introduces {additional} challenges. 
For ease of interaction, we expect sketches specified only for a sparse set of keyframes, making it difficult to generate frames without sketch inputs.
Additionally, video generation costs significantly higher memory resources, making methods \cite{ControlNet, PIXART_delta} that replicate half of the base model as a sketch encoder {easily out of memory.}}

\textbf{Diffusion-based Video Generation.}
VDM~\cite{VDM} pioneered diffusion-based video generation with a 3D U-Net denoising network.
To improve quality, subsequent works~\cite{Text2Video-Zero, Align_Your_Latents, AnimateDiff, LVDM, ModelScope, SVD, VideoCrafter1, VideoCrafter2} integrated temporal modules into text-to-image models~\cite{StableDiffusion} to enable text- and image-conditioned video synthesis.
Considering the efficiency, the DiT architecture is further used in Sora \cite{sora} and open-source projects~\cite{opensora, Open-Sora-Plan, Vchitect}.
Despite these advancements, subtle flickering artifacts remain in the %
results.
CogVideoX~\cite{cogvideo, cogvideox} further proposes a 3D full attention that merges the spatial and temporal attention, facilitating the generation of long-duration and high-resolution videos.

\begin{figure*}[h]
    \centering
    \includegraphics[width=1.0\linewidth]{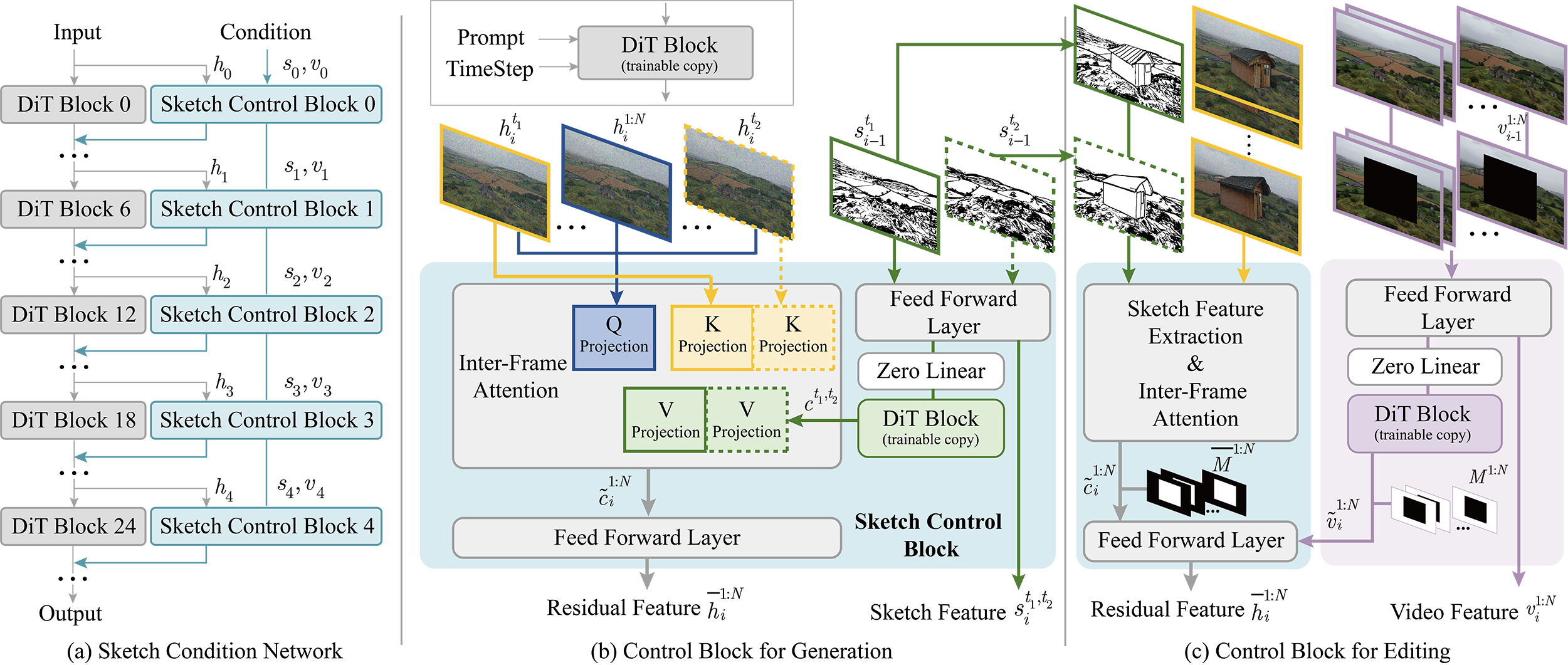}
    \caption{Our framework for sketch-based video generation and editing. 
    (a) Our sketch condition network for the DiT-based video generation architecture has a skip structure and five sketch control blocks that predict residual features. 
    (b) For generation, features are extracted from temporally sparse input sketches and propagated through inter-frame attention. The input sketches are provided for one or two keyframes (the second sketch is shown as a dotted line). 
    {In the top left corner {of} (b), the prompt and timestep inputs are shown.}
    {(c)} For editing, {the same sketch control block (b) is utilized, with an additional video insertion module and video masks $M^{1:N}$ to analyze} 
    the relationship between edited and unedited regions. 
    The 3D causal VAE is omitted to save space.
    }
    \label{fig:pipeline}
\end{figure*}

Building on existing video generation models, various conditions have been introduced to control the generation,
such as camera movement~\cite{Direct-a-Video, MotionCtrl, CameraCtrl}, subject identity~\cite{CustomCrafter, VideoBooth}, key-point trajectory~\cite{Image_Conductor, Motion-I2V, Tora} and example motions~\cite{MotionBooth, MotionCrafter}.
However, they often lack control over spatial layouts and geometric details.
Some methods~\cite{ControlVideo, Control-A-Video, Video_ControlNet} address this by extending image-based ControlNet~\cite{ControlNet} to videos but require {all-frame} conditions that are tedious for sketch interaction.
SparseCtrl~\cite{SparseCtrl} tackles this by using white images for completions. 
However, due to the restriction of its base model~\cite{AnimateDiff} and a simple network design, its results suffer from temporal flickering. 
Similar ideas have been applied to cartoon interpolation~\cite{ToonCrafter} and colorization~\cite{LVCD} with line art as input, but their outputs are limited to cartoon style.
Our method utilizes sparse inputs, including hand-drawn sketches on one or two keyframes, to generate temporally stable and realistic videos. 
Additionally, our method further supports the sketch-based detailed editing of existing videos.

\textbf{Deep Learning-based Video Editing.}
{Pioneer works achieve video editing by style transfer~\cite{Style_Transfer_video, Stylize_Video_example}, GAN inversion~\cite{Stitch_it_in_Time, DFVD}, and layered representations~\cite{Layered_Video_Editing}.}
The advent of diffusion models further provides new editing paradigms.
One category of such methods leverages image generation models to achieve compelling editing results, using temporal consistency techniques such as layered representations~\cite{StableVideo, CoDeF}, cross-frame attention~\cite{EVE, Pix2Video, InsV2V, slicedit} and pixel warping~\cite{Rerender_A_Video}. 
A second category of %
methods employ video generation models. %
These works propagate the edits applied on the first frame into the other frames %
\cite{I2VEdit, AnyV2V}, or utilize an inpainting strategy to achieve text-based editing~\cite{AVID} {and} motion modification~\cite{ReVideo}. 
{To achieve text-based large-scale shape modification while maintaining motion features, a space-time feature loss \cite{space_time} is designed to guide the inference process.
Our method moves beyond traditional text- or image-based editing approaches, allowing users to draw one or two keyframe sketches at arbitrary time points for interactive video editing.}
Additionally, our method effectively handles sketch-based shape manipulation and dynamic object insertion, which are challenging for previous works. 

\textbf{Controllable Attention Mechanism.}
Attention across frames is initially used to ensure temporal consistency in AnimateDiff~\cite{AnimateDiff} and subsequent %
video generation models.
Subsequent works, such as VideoBooth~\cite{VideoBooth} and Still-Moving~\cite{Still_Moving}, utilize it to learn identity-aware features for customized video generation.
For video editing, existing works~\cite{TokenFlow, InsV2V, EVE} utilize cross-frame attention to capture the temporal motion of input videos, enabling effective propagation of editing operations.
Our method %
applies this idea to pixel-aligned sketch-based video generation and propagates the spatial geometric conditions instead of identity customization. 
We use a new feature derivation strategy for enhanced spatial control, differentiating our approach from traditional cross-attention mechanisms.

\section{Methodology}

This section introduces our sketch-based video generation and editing framework.
In Sec.\ref{sec:preliminary}, we provide an overview of CogVideoX-2b~\cite{cogvideox}, a pretrained text-to-video generation method, which we will use for sketch-based video generation in our work.
In Sec.\ref{sec:Generation}, we describe our sketch condition network specifically designed for the DiT architecture, which contains sketch control blocks to predict residual features.
Within each control block, an inter-frame attention mechanism is designed to propagate the {input} sketches.
In Sec.\ref{sec:Editing}, we detail the editing framework, which incorporates a video insertion module and latent fusion to preserve the features of the original video. %

\subsection{Preliminary}\label{sec:preliminary}

CogVideoX~\cite{cogvideox} is a text-to-video generation model that employs a 3D {causal VAE} and builds diffusion within the latent space.
In CogVideoX-2b, the VAE model performs 8×8 spatial and 4× temporal downsampling, followed by a DiT architecture with 30 blocks for video generation. 
Within each block, a 3D full attention processes the concatenated text embeddings and patchified video latents, followed by a feed-forward layer to output the features. 
The 3D full attention merges the commonly used separate spatial and temporal attention~\cite{AnimateDiff, SVD, opensora} to improve the temporal coherence. 
The training objective of diffusion is:
{\begin{minipage}{\linewidth}
\begin{footnotesize} 
\begin{equation}
    L(\theta):={\rm{E}}_{t,x_0^{1:N},y,\epsilon} \left \| \epsilon - \epsilon_{\theta} (\sqrt{\bar{\alpha} _{t} } x_0^{1:N} + \sqrt{1 - \bar{\alpha}_{t} } \epsilon ,t, y) \right \| ^ 2,
\end{equation}
\end{footnotesize}
\end{minipage}}
where $t$ is sampled between 1 and $\rm{T}$ {(denosing steps)}, $\epsilon$ is the random noise, $y$ is text prompts, and $x_0^{1:N}$ is the video data with $N$ frames. 
The v-prediction~\cite{v_prediction} and zero SNR~\cite{zero_SNR} are utilized for the diffusion setting. 

\subsection{Sketch-based Video Generation}\label{sec:Generation}
Given text prompts and one or two keyframe sketches with corresponding time points {$t_1, t_2$}, our method generates video clips that respect the input text prompts and sketches.
As shown in Fig. \ref{fig:pipeline}, we design a sketch condition network for effective control.
The input sketches are encoded into the latent space by the pretrained VAE, followed by patchifying and time-aware positional embedding to generate sketch latents $s_0^{t_1,t_2}$. 

\textbf{Skip Residual Structure.} 
In sketch-based image generation, methods like ControlNet \cite{ControlNet} and PIXART-$\delta$~\cite{PIXART_delta} copy the half of the base model as their sketch encoder to fully utilize the pretrained text-to-image models.
However, applying this to video generation, as done in SparseCtrl~\cite{SparseCtrl}, is memory inefficient %
because it requires adding half of the base model's parameters. 
Unlike the U-Net architecture, the DiT network does not have an explicit encoder and decoder. 
Therefore, the assumption in PIXART-$\delta$~\cite{PIXART_delta} that the first half of blocks serve as the encoder can be improved.

\begin{figure*}[ht]
    \centering
    \includegraphics[width=1.0\linewidth]{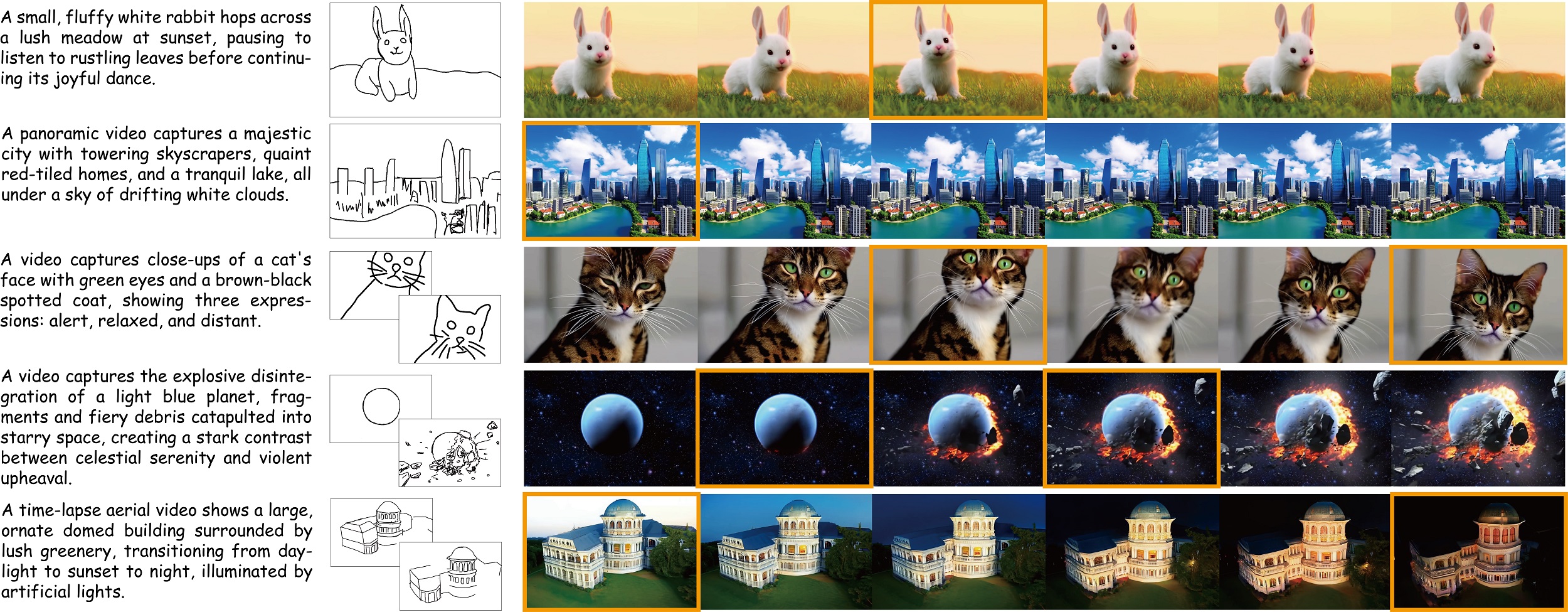}
    \caption{The sketch-based video generation results.
    Left: The input text prompts and sketches.  %
    Right: %
    The video generation results. %
    It can be seen that the generated results show high quality and good faithfulness with the input sketches. 
    Our method can handle one/two keyframe sketch(es) at arbitrary user-specified time points {(the frames corresponding to the input time points are highlighted by orange)}. 
    }
    \label{fig:result_gen}
\end{figure*}

Instead of borrowing local consecutive blocks of the base model as the encoder and predicting their residual features, we recognize that blocks at different depths process distinct feature levels, which should be considered for sketch control.
As illustrated in Fig.~\ref{fig:pipeline} (a), we propose a novel skip residual structure that reduces the number of blocks while enabling effective sketch control and high-quality generation. 
Our sketch condition network contains %
5 {%
sketch control blocks,} %
uniformly distributed across the pretrained generation network to predict residual features for blocks 0, 6, 12, 18, and 24 of the original video generation model.
This structure efficiently integrates sketch control information into multiple feature levels, enhancing the analysis of input conditions and original semantic features.

\textbf{Sketch Control Block.}
In the $i$-th sketch control block, the input consists of hidden video features $h_{i}^{1:N}$ and sketch features $s_{i-1}^{t_1,t_2}$, and the output is the residual features $\overline{h}_{i}^{1:N}$ and updated sketch features $s_{i}^{t_1,t_2}$. 
As shown in Fig. \ref{fig:pipeline} (b), the sketch feature $s_{i-1}^{t_1,t_2}$ is processed: 
\begin{equation}
s_{i}^{t_1,t_2}={\rm{FeedForward}}(s_{i-1}^{t_1,t_2}),
\end{equation}
where the output sketch feature $s_{i}^{t_1,t_2}$ is used for sketch control propagation and as the input to the next sketch control block.

To propagate the sketch inputs, a direct approach~\cite{SparseCtrl} replaces missing sketches with white images and employs a trainable copy of the pretrained DiT block to predict residual features. 
However, as shown in Fig.~\ref{fig:compare_gen}, this leads to fuzzy details in challenging cases, as the network randomly processes both sketches and white images. %
{This mixed input} is far from the pretrained model's task.
To address the above issue, we employ {a} trainable copy of {pretrained} DiT block to process only the sketch inputs (no white placeholder), 
aligning with the pretrained weights and reducing learning difficulty. 
The resulting keyframe {sketch} features, denoted as $c^{t_1,t_2}_{i}$, are propagated to all frames {through an} inter-frame attention approach. 

\textbf{Inter-frame Attention.}
We utilize the input hidden features of all frames to calculate $Q$ and the hidden features {corresponding to} the control frames to calculate $K$. 
During attention computation, this approach captures the internal relationship between all frames and control keyframes, allowing propagation of $V$ (derived from the keyframe
sketch features).
Our inter-frame attention is distinct from typical cross-frame attention, which uses both $K$ and $V$ from control conditions; instead, we leverage the %
{inter-frame similarity} within the input noisy hidden video features and progressively insert the sketches' spatial and temporal information. %
The output $\widetilde{c}^{1:N}_{i}$ is computed as: ${\rm{Attention}}(Q,K,V)={\rm{Softmax}}(\frac{QK^T}{\sqrt{d} }) \cdot V$, with
\begin{equation}
{\rm{Q}}=W_q \cdot h^{1:N}_{i}, K=W_k \cdot h^{t_1,t_2}_{i}, V=W_v \cdot c^{t_1, t_2}_{i}, 
\end{equation}
where $W_q, W_k, W_v$ are trainable linear projection weights. 
The output of the inter-frame attention is fed into a feed-forward layer to generate final residual features $\overline{h}_{i}^{1:N}$.

\textbf{Training Strategy.} 
To train the sketch condition network, we employ a hybrid training strategy in two stages. 
In the first stage, to accelerate convergence and address the issue %
of limited video data, the network is trained both for image generation at arbitrary time points and video generation with one or two keyframe sketches {(randomly selected from corresponding sketch videos).} %
In the second stage, video data alone is used to improve the temporal coherence. 

\begin{figure*}[ht]
    \centering
    \includegraphics[width=1.0\linewidth]{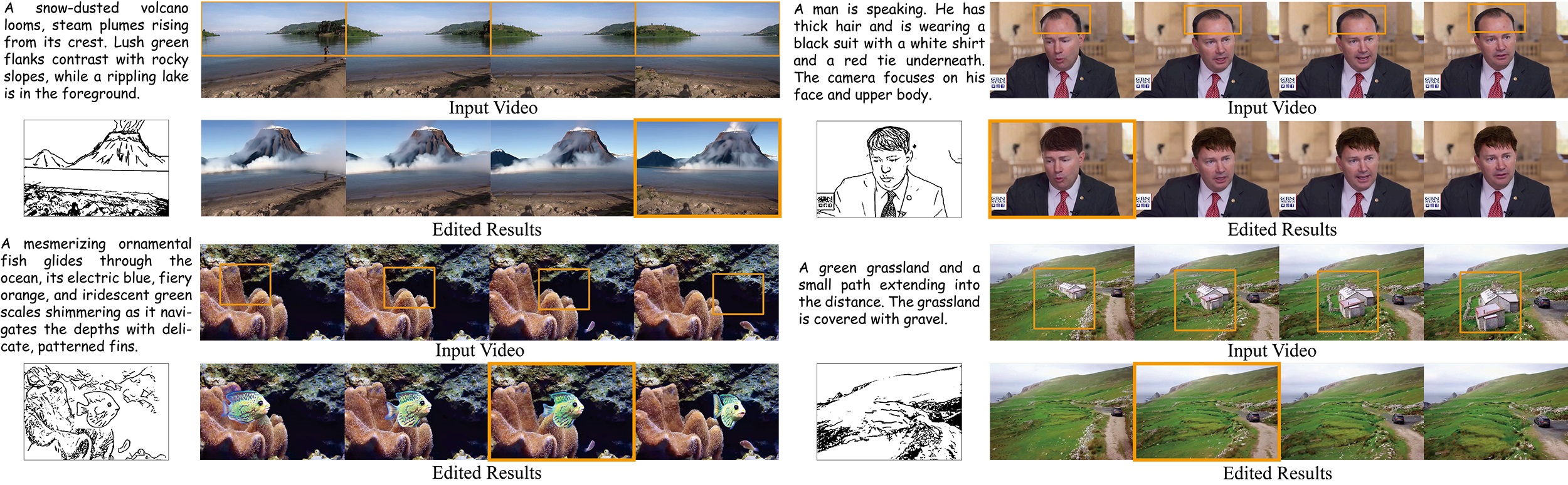}
    \caption{The sketch-based video editing results.
    For each example, the text prompts and sketches are shown on the left.
    On the right, the input real videos are shown at the top, while the edited results with the control keyframe highlighted in orange are shown at the bottom.
    The editing region {masks {are} manually provided by users,} {highlighted} as orange boxes. 
    Our method generates realistic local editing results. 
    }
    \label{fig:result_editing}
\end{figure*}

\subsection{Sketch-based Video Editing}\label{sec:Editing}
{For editing, given real or synthetic videos, users select {one or two} keyframes at arbitrary time points and modify the extracted sketches, with additional inputs of text prompts and masks $M^{1:N}$ that label regions {to be edited} for all the frames. 
Our method then generates realistic, local editing results. 
The input video is multiplied by the inverted masks to remove information from the edited regions and is subsequently encoded to generate a masked video latent representation $v^{1:N}_0$. 
}

\textbf{Video Insertion Module.}
For sketch-based editing, newly generated contents within the mask regions should be coherent with the original spatial and temporal features in the unedited regions. 
Thus, we design a video insertion module that analyzes the relationship between the input sketches and the original video.
The video insertion module takes $v^{1:N}_{i-1}$ {(input video latent or generated by a previous control block)} %
as input and predicts the updated video features $v^{1:N}_{i}$, similar to the sketch generation process. 
Since video features are not temporally sparse, we use a trainable copy of CogVideoX-2b's DiT block to directly generate video insertion features $\widetilde{v}^{1:N}_{i}$. 
The sketch branch output $\widetilde{c}^{1:N}_{i}$ and video branch output $\widetilde{v}^{1:N}_{i}$ are multiplied by their respective masks and concatenated: 
\begin{equation}
{\rm{Concat}}(\widetilde{c}^{1:N}_{i} * M^{1:N}, \widetilde{v}^{1:N}_{i} * \overline{M}^{1:N}),
\end{equation}
which serves as inputs to the feed-forward layers, producing final residual features incorporating the original video and sketch control information. 
This design ensures seamless integration of new contents with the original videos, effectively propagating edits across frames for dynamic motion.

\textbf{Training Strategy.}
Directly training the video editing network would lead to low fidelity with the input sketches, possibly because of the challenging interaction between the input sketches and videos. 
So we finetune it from the pretrained sketch condition network for generation and add the new video insertion module. 
The pretrained model already has good sketch fidelity and only requires learning video information. 
The network is trained in a self-supervised inpainting manner, with randomly generated masks to imitate real-world editing.  

\textbf{Inference Latent Fusion.}
Although the original videos are encoded in the condition network, as shown in Fig.~\ref{fig:ablation_study_editing}, fine details might be lost during editing. 
To address this, we propose a latent fusion approach at inference. 
Specifically, we apply DDIM inversion~\cite{DDIM} to generate noisy latent codes of input videos across the inference steps. 
At Steps 25 and 49 (out of 50 total steps), the latent codes in the unedited regions are replaced with these inversion latent codes, ensuring better preservation of the original video’s details and improving the coherence of the edited regions.

\section{Evaluation}

\subsection{Implementation Details}\label{sec:implement_details}

We implement {\sysName} based on CogVideoX-2b~\cite{cogvideox}, trained on a subset of OpenVid~\cite{OpenVid_1M} and LAION~\cite{LAION_5B} datasets, with paired sketches from \cite{Line_drawing}. 
Training uses 8 NVIDIA H800 GPUs with a batch size of 8 and gradient accumulation of 4.
For generation, one or two keyframe sketches are randomly sampled from the video, %
with 10,000 steps for each training stage. 
For editing, randomly drawn masks are applied and trained for 20,000 steps. 
For ease of reading, the input text prompts are simplified in figures. 
Additional implementation details and full input texts are available in the supplementary material.

\begin{figure}[h]
    \centering
    \includegraphics[width=1.0\linewidth]{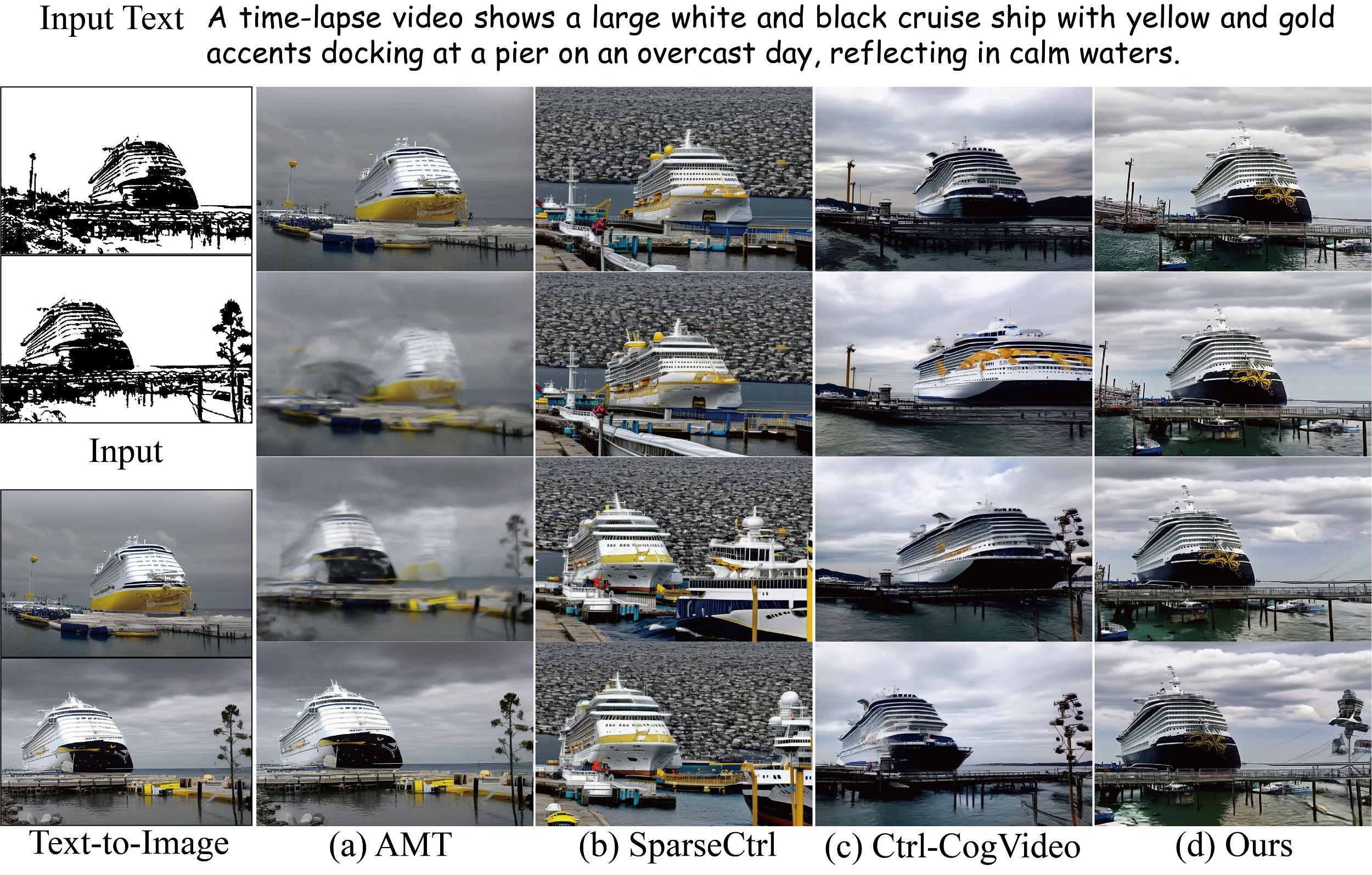}
    \caption{The comparison results of sketch-based video generation.
    {Text prompts are shown on the top.}
    On the left, we show the input sketches and sketch-based image generation results by ControlNet~\cite{ControlNet}.
    On the right, we show the results of the compared approaches, including AMT~\cite{AMT}, SparseCtrl~\cite{SparseCtrl}, Ctrl-CogVideo~\cite{ControlNet}, and ours. 
    Our method produces better results, especially for the intermediate frames. 
    }
    \label{fig:compare_gen}
\end{figure}

\begin{figure}[h]
    \centering
    \includegraphics[width=1.0\linewidth]{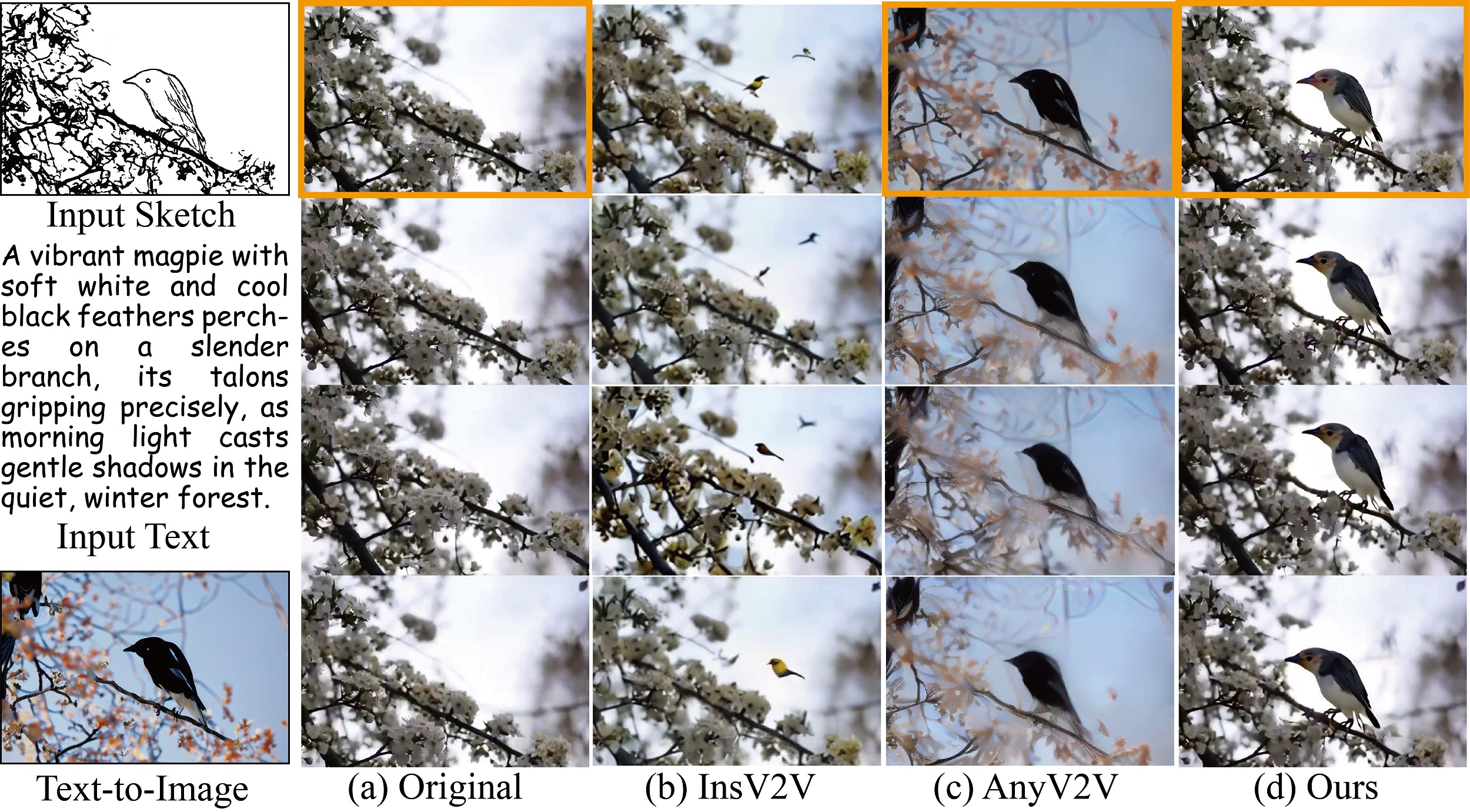}
    \caption{The comparison results of sketch-based video editing. On the left, we show the drawn editing sketches {(for the frames highlighted in orange)}, text prompts, and sketch-based image generation results by ControlNet~\cite{ControlNet}. 
    On the right, we show the original videos and editing results by the compared methods, including InsV2V~\cite{InsV2V}, AnyV2V~\cite{AnyV2V}, and ours.
    Our method generates the most realistic results and preserves unedited regions well.
    }
    \vspace{-1mm}
    \label{fig:compare_editing}
\end{figure}

\subsection{Results}\label{sec:Results}

Our method generates high-quality videos from one or two keyframe sketches and text prompts. 
As shown in Fig.~\ref{fig:result_gen}, our method can accurately control the object shape and scene layout, such as the rabbit's pose (1st row) and the position of lakes and buildings (2nd row).
Text-only inputs cannot achieve such detailed geometry control.
{Our method also achieves interesting dynamic motion interpolation and extrapolation, generating smooth and realistic transitions, as seen in the cat’s head shaking {(3rd row}) and the building's movement (last row).}
This allows control over both spatial layout and dynamic motion.

Our method also supports sketch-based video editing. 
Users specify bounding boxes and draw sketches in {those} %
regions, %
and our method generates photorealistic, seamlessly integrated content. 
As shown in Fig.~\ref{fig:result_editing}, the new contents blend well with unedited regions and have dynamic motions like hat/hair rotation with original objects (1st row) or interesting fish swimming (2nd row). 
Our method can handle diverse editing cases, including object insertion, component replacement, and object removal. 
The unedited regions are well-preserved thanks to our latent fusion approach.

\subsection{Comparison}\label{sec:Comparison}
For sketch-based generation, we compare our methods with three methods {given the same} %
{keyframe sketches {text prompts as}} inputs. For SparseCtrl~\cite{SparseCtrl}, we use the official pretrained model and extract sketches by HED~\cite{HED}. 
{It is trained on videos in the WebVid-10M dataset \cite{WebVid-10M} instead of the OpenVid-1M dataset \cite{OpenVid_1M} used in our method.}
We extend SparseCtrl to CogVideoX-2b~\cite{cogvideox}, using PIXART-{\(\delta\)}~\cite{PIXART_delta} as the sketch condition encoder (with 5 DiT blocks same as our method) and white images to complete the missing condition frames. 
We also compare with an interpolation baseline, which uses ControlNet to translate two-frame sketches into images and then interpolates them with AMT~\cite{AMT}.
For the sketch-base editing task, we compare ours with a text-based video editing method InsV2V~\cite{InsV2V} and a first-frame editing method AnyV2V~\cite{AnyV2V}.
{We compare with additional methods \cite{ToonCrafter, Seine, I2VEdit, TokenFlow}, as shown in supplemental material.}

As shown in Fig.~\ref{fig:compare_gen}, ControlNet~\cite{ControlNet} translates sketches into realistic images{, which, however,} lack temporal consistency and vary with shading and content. 
The interpolation results of AMT~\cite{AMT} exhibit %
fuzzy details and artifacts. 
SparseCtrl~\cite{SparseCtrl}, based on AnimateDiff~\cite{AnimateDiff}, exhibits temporal flickering, such as the suddenly appearing tree and distorted tower top ({see the orange boxes in the} 2nd row).
Extending SparseCtrl to CogVideoX-2b~\cite{cogvideox} (Ctrl-CogVideo) still generates fuzzy details in intermediate frames, possibly due to the CogVideoX-2b's pretrained self-attention being designed for dense inputs instead of sparse sketches. 
Our method generates realistic videos with clear details and good temporal coherence, with even small details like the electric wires in the top-right corner accurately propagated. 

For sketch-based editing (Fig.~\ref{fig:compare_editing}), InsV2V~\cite{InsV2V} generates interesting results with birds but lacks control over shape and geometry through text prompts alone.
{For image-based video editing method}, we utilize ControlNet {to edit the first frame} and then propagate editing into the video by AnyV2V~\cite{AnyV2V}. 
However, since the motion is borrowed from the original video, AnyV2V struggles with new content, leading to fuzzy details and distortion in the bird, as well as changes in unedited regions due to ControlNet’s inability to retain the original features. In contrast, 
our method produces more realistic results with faithful representations of the sketch. 
More generation and editing results are available in the supplemental material and video. 

\begin{table}\scriptsize %
  \centering
  \begin{tabular}{cccccc}
    \toprule
    \makebox[0.05\textwidth][c]{Methods} & 
    \makebox[0.05\textwidth][c]{LPIPS $\downarrow$} & \makebox[0.05\textwidth][c]{CLIP $\uparrow$} & \makebox[0.05\textwidth][c]{Fidelity} & \makebox[0.05\textwidth][c]{Consistency} & \makebox[0.05\textwidth][c]{Realism} \\ 
    \midrule
    {AMT} & 29.17 & 96.12 & 3.13 & 3.51 & 3.57 \\
    {SparseCtrl} & 44.85 & 96.48 & 2.79 & 2.94 & 2.83 \\
    {Ctrl-CogVideo} & 32.23 & 98.04 & 2.86 & 2.47 & 2.50 \\
    {Ours} & \cellcolor{orange}{27.56} & \cellcolor{orange}{98.31} & \cellcolor{orange}{1.21} & \cellcolor{orange}{1.08} & \cellcolor{orange}{1.11} \\
    \bottomrule
  \end{tabular}
  \caption{The quantitative results of sketch-based video generation comparison. The LPIPS and CLIP numbers are scaled up 100×, with each cell colored to indicate the \colorbox{orange}{best}.}
  \vspace{-2mm}
  \label{tab:comp_gen}
\end{table}

\begin{table}\scriptsize
  \centering
  \begin{tabular}{ccccccc}
    \toprule
    \makebox[0.04\textwidth][c]{Methods} & 
    \makebox[0.04\textwidth][c]{LPIPS $\downarrow$} & \makebox[0.04\textwidth][c]{CLIP $\uparrow$} & \makebox[0.04\textwidth][c]{PSNR$\uparrow$}
    & \makebox[0.04\textwidth][c]{Fidelity} & \makebox[0.04\textwidth][c]{Preservation} & \makebox[0.04\textwidth][c]{Realism} \\ 
    \midrule
    {InsV2V} & 13.61 & 95.39 & 16.84 & 2.58 & 2.26 & 2.61 \\
    {AnyV2V} & 11.92 & 93.47 & 13.68 & 2.35 & 2.69 & 2.34\\
    {Ours} & \cellcolor{orange}{9.74} & \cellcolor{orange}{98.34} & \cellcolor{orange}{36.48} & \cellcolor{orange}{1.07} & \cellcolor{orange}{1.05} & \cellcolor{orange}{1.04} \\
    \bottomrule
  \end{tabular}
  \caption{The quantitative results of sketch-based video editing comparison. The LPIPS and CLIP numbers are scaled up 100×, with each cell colored to indicate the \colorbox{orange}{best}.}
  \label{tab:comp_editing}
\end{table}

We follow SparseCtrl~\cite{SparseCtrl} and use the LPIPS~\cite{LPIPS} metric to measure sketch faithfulness between the input sketches and those extracted from the corresponding frames of the generated videos. 
We use the CLIP~\cite{CLIP} similarity to assess temporal coherence. 
For sketch-based generation, we test 200 random examples from OpenVid~\cite{OpenVid_1M}, using the first and last frame sketches {and the corresponding text prompts in the dataset.} %
As shown in Table~\ref{tab:comp_gen}, our method achieves the lowest LPIPS and highest CLIP scores, indicating its superior performance. 
For sketch-based editing, we use additional an MSE metric to measure unedited region preservation. 
We utilize 10 examples with hand-drawn sketches as input. 
In Table~\ref{tab:comp_editing}, our method outperforms existing methods across all metrics, demonstrating its superiority. 

\begin{figure}[h]
    \centering
    \includegraphics[width=1.0\linewidth]{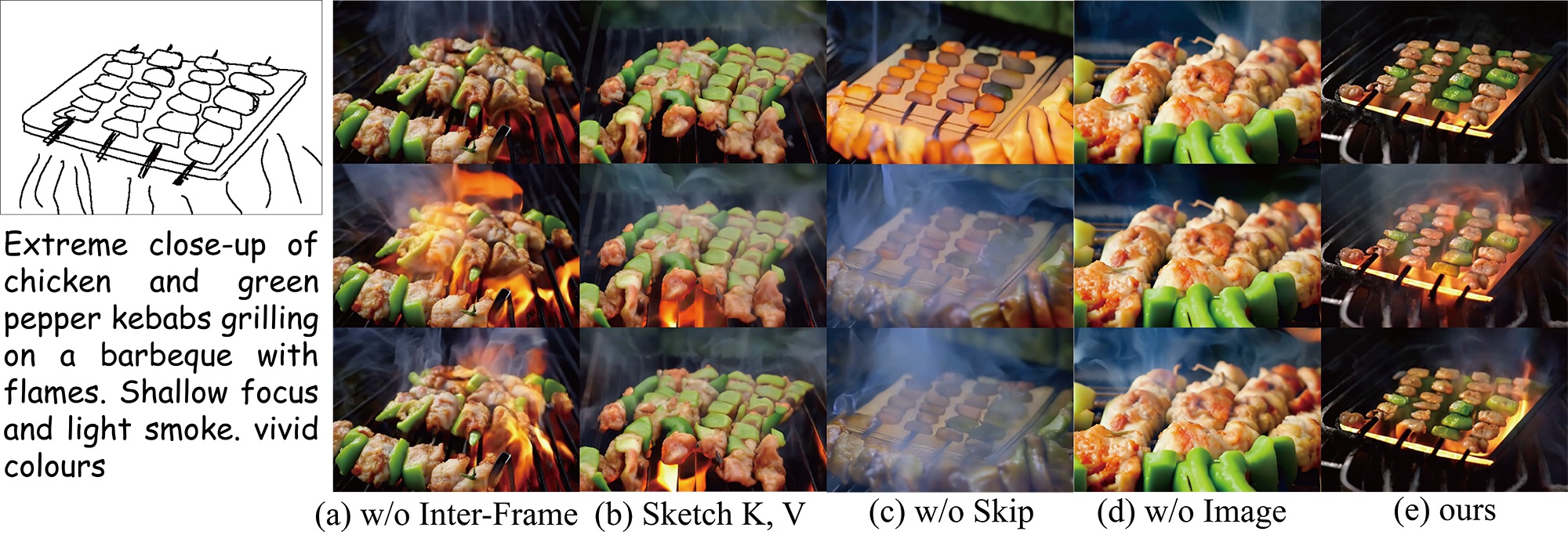}
    \caption{The ablation study results of sketch-based video generation. Our method generates more realistic and sketch-faithful results than the baselines. 
    }
    \label{fig:ablation_study_gen}
\end{figure}

\begin{figure} %
    \centering
    \includegraphics[width=1\linewidth]{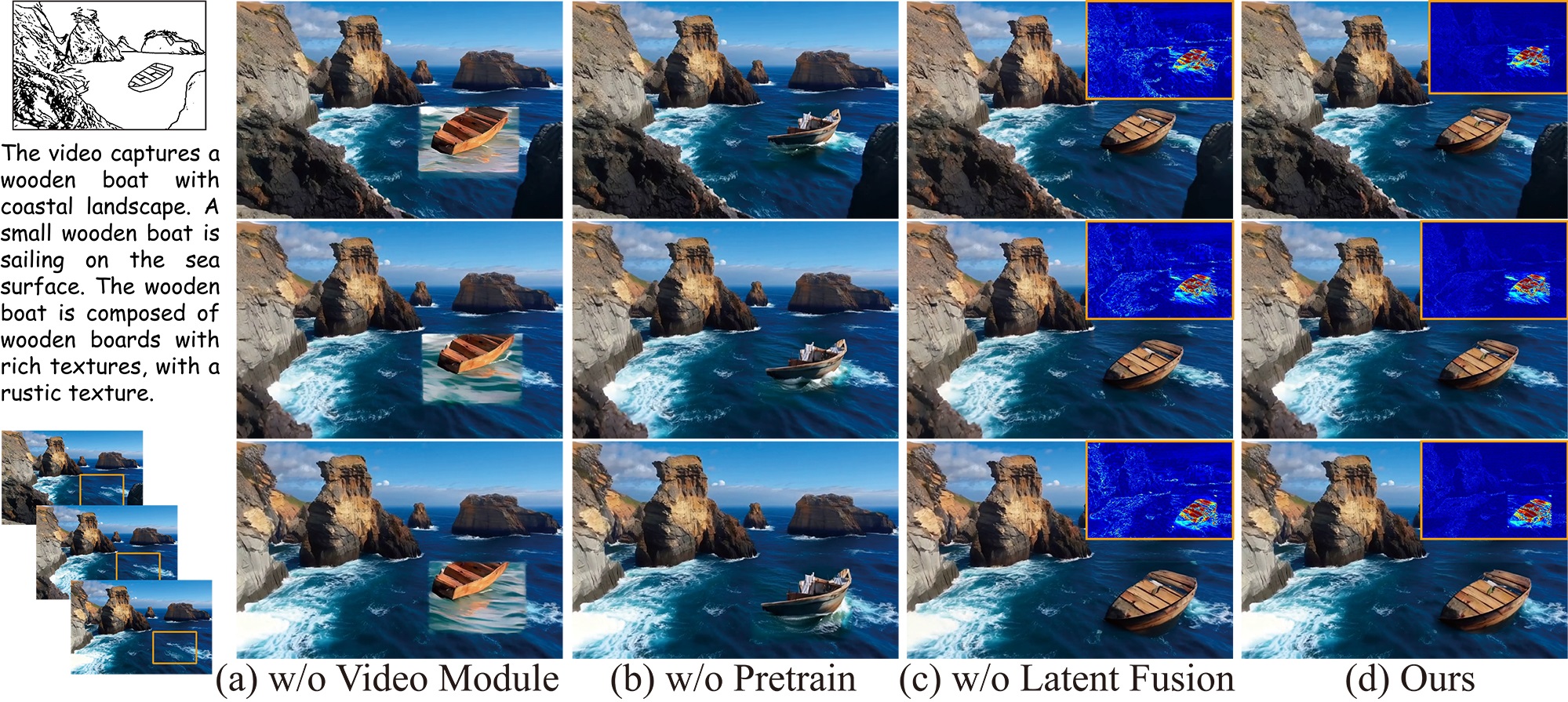}
    \vspace{-5mm}
    \caption{The ablation study results of sketch-based video editing. Our method generates realistic results faithful to sketches. 
    We utilize heat maps (top right) to visualize the difference between the edited and original frames. 
    }
    \vspace{-2mm}
    \label{fig:ablation_study_editing}
\end{figure}

\subsection{Ablation Study}\label{sec:Ablation}

We conduct an ablation study to evaluate the effectiveness of each key component. 
Due to the restriction of computing resources, all the models are trained with a batch size of 2.
For sketch-based generation, removing inter-frame attention and concatenating sketches with frame features at the temporal dimension results in strange control with low sketch faithfulness (Fig.~\ref{fig:ablation_study_gen} (a)). 
Similarly, if we replace the inter-frame attention with a typical cross-attention that uses sketch features for both Key and Value (b), the sketch fidelity is also degraded due to the lack of effective control propagation. 
If we remove the skip structure and predict residual features for the first 5 blocks (c), the realism of the generated results is negatively influenced due to less control ability. 
Removing image data during training (d) causes significant geometry mismatch with input sketches. 
Compared to these alternatives, our method produces the most realistic and faithful results. As shown in Table~\ref{tab:ablation_gen}, our method has the lowest LPIPS values (supporting the best sketch fidelity) while the CLIP metric has similar results (human eyes cannot recognize temporal coherence difference).

In sketch-based video editing (Fig.~\ref{fig:ablation_study_editing}), removing the video insertion module (a) and solely utilizing the sketch condition encoder results in good sketch fidelity but obvious inconsistency with unedited regions. 
Training the sketch condition encoder from scratch without generation pretraining (b) reduces the faithfulness to the input sketches in the edited regions, which is also validated by the high LPIPS in Table \ref{tab:ablation_gen}.
We use heat maps to reveal differences between edited and original frames. 
Removing the latent fusion strategy changes unedited region details (c), such as mountains and waves, as shown in the heat maps. 
The low PSNR value in unedited regions also supports this, as shown in Table \ref{tab:ablation_gen}.
Our method maintains strong sketch fidelity, consistency, and unedited region preservation.

\begin{table}\footnotesize
  \centering
  \begin{tabular}{cccccc}
    \toprule
    \makebox[0.05\textwidth][c]{Metric} & \makebox[0.08\textwidth][c]{w/o Inter-Frame} & \makebox[0.06\textwidth][c]{Sketch K,V} & \makebox[0.05\textwidth][c]{w/o Skip} & \makebox[0.05\textwidth][c]{w/o Image} & \makebox[0.04\textwidth][c]{Ours}\\ 
    \midrule
    {LPIPS $\downarrow$} & 36.33 & 32.59 & \cellcolor{yellow}31.91 & 34.58 & \cellcolor{orange}{30.79} \\
    {CLIP $\uparrow$} & 98.10 & \cellcolor{yellow}98.19 & 97.60 & 98.24 & \cellcolor{orange}98.48 \\
    \bottomrule
  \end{tabular}
  \begin{tabular}{ccccc}
      \makebox[0.05\textwidth][c]{Metric} & \makebox[0.09\textwidth][c]{w/o Video module} & \makebox[0.08\textwidth][c]{w/o Pretrain} & \makebox[0.09\textwidth][c]{w/o Latent Fusion} & \makebox[0.05\textwidth][c]{Ours} \\ 
      \midrule
    {LPIPS $\downarrow$} & \cellcolor{orange}{9.31} & 12.60 & 9.77 & \cellcolor{yellow}9.74 \\
    {CLIP $\uparrow$} & 97.97 & 98.12 & \cellcolor{orange}{98.44} & \cellcolor{yellow}98.34 \\
    {PSNR $\uparrow$} & \cellcolor{yellow}33.61 & 36.05 & 31.69 & \cellcolor{orange}{36.48} \\
    \bottomrule
    \end{tabular}
  \caption{The qualitative results of ablation study for sketch-based generation and editing, with each cell colored to indicate the \colorbox{orange}{best} and \colorbox{yellow}{second best}.}
  \label{tab:ablation_gen}
\end{table}

\subsection{User Study}\label{sec:Perceptual_Study}

We conducted a user study to validate that our method outperforms existing approaches. 
For video generation, we randomly selected 10 examples from the test set in Sec.~\ref{sec:Comparison}. 
In the questionnaire, participants were shown the results of different methods in random order. 
20 participants ranked the methods results by three criteria: Sketch Fidelity, Temporal Consistency, and Video Realism. 
The rankings provided by the participants were used as scores. 
As shown in Table~\ref{tab:comp_gen}, our method outperforms existing methods in all criteria, demonstrating its superior performance. 
For sketch-based editing, we used the same 10 examples from the test set in Sec.~\ref{sec:Comparison}.
The same participants ranked the methods based on three criteria: Sketch Fidelity, Unedited Region Preservation, and Video Realism. 
As shown in Table~\ref{tab:comp_editing}, our method also outperforms existing approaches in editing.

\section{Conclusion and Discussion}

We presented {\sysName}, a unified method for sketch-based video generation and editing. 
For generation, we proposed a sketch condition network that predicts residual features for skipped DiT blocks of the base models to save the memory and achieve effective control.
An inter-frame attention is further proposed to propagate the keyframe sketches, achieving interesting motion interpolation or extrapolation. We also introduce a hybrid image and video training strategy. 
For editing, we incorporate a video insertion module to ensure the newly generated content is spatially and temporally coherent with the original video. 
During inference, a latent fusion approach preserves unedited regions accurately. 

\textbf{Limitation and Future Work.}
While our method generates high-quality videos, its capabilities are still limited by the pretrained text-to-video model.
Enhancing performance with more powerful pretrained models and generating long videos instead of video clips is a potential avenue. 
Additionally, similar to image generation, our method struggles with too complex scenarios, such as human hands and interaction between multiple objects. 
Incorporating 3D priors~\cite{NeRF_survey, 3D_survey, 3DGS_survey} like SMPL-X~\cite{SMPL-X:2019} helps address these issues in human scenarios. 
Moreover, while our method focuses on geometry control, exploring appearance customization through tools like color strokes presents an intriguing area for future research.

\section{Acknowledgments}
This work was sponsored by CCF-Kuaishou Large Model Explorer Fund, Beijing Municipal Science and Technology Commission (No. Z231100005923031), Innovation Funding of ICT, CAS (No. E461020) and National Natural Science Foundation of China (No. 62322210).
The authors would like to acknowledge the Nanjing Institute of InforSuperBahn OneAiNexus for providing the training and evaluation platform.

{
    \small
    \bibliographystyle{ieeenat_fullname}
    \bibliography{main}

\begin{thebibliography}{86}
\providecommand{\natexlab}[1]{#1}
\providecommand{\url}[1]{\texttt{#1}}
\expandafter\ifx\csname urlstyle\endcsname\relax
  \providecommand{\doi}[1]{doi: #1}\else
  \providecommand{\doi}{doi: \begingroup \urlstyle{rm}\Url}\fi

\bibitem[Bain et~al.(2021)Bain, Nagrani, Varol, and Zisserman]{WebVid-10M}
Max Bain, Arsha Nagrani, G{\"{u}}l Varol, and Andrew Zisserman.
\newblock Frozen in time: {A} joint video and image encoder for end-to-end retrieval.
\newblock In \emph{Int. Conf. Comput. Vis.}, pages 1708--1718, 2021.

\bibitem[Blattmann et~al.(2023{\natexlab{a}})Blattmann, Dockhorn, Kulal, Mendelevitch, Kilian, Lorenz, Levi, English, Voleti, Letts, Jampani, and Rombach]{SVD}
Andreas Blattmann, Tim Dockhorn, Sumith Kulal, Daniel Mendelevitch, Maciej Kilian, Dominik Lorenz, Yam Levi, Zion English, Vikram Voleti, Adam Letts, Varun Jampani, and Robin Rombach.
\newblock Stable video diffusion: Scaling latent video diffusion models to large datasets.
\newblock \emph{CoRR}, abs/2311.15127, 2023{\natexlab{a}}.

\bibitem[Blattmann et~al.(2023{\natexlab{b}})Blattmann, Rombach, Ling, Dockhorn, Kim, Fidler, and Kreis]{Align_Your_Latents}
Andreas Blattmann, Robin Rombach, Huan Ling, Tim Dockhorn, Seung~Wook Kim, Sanja Fidler, and Karsten Kreis.
\newblock Align your latents: High-resolution video synthesis with latent diffusion models.
\newblock In \emph{IEEE Conf. Comput. Vis. Pattern Recog.}, pages 22563--22575, 2023{\natexlab{b}}.

\bibitem[Byeon et~al.(2022)Byeon, Park, Kim, Lee, Baek, and Kim]{COYO_700M}
Minwoo Byeon, Beomhee Park, Haecheon Kim, Sungjun Lee, Woonhyuk Baek, and Saehoon Kim.
\newblock Coyo-700m: Image-text pair dataset.
\newblock \url{https://github.com/kakaobrain/coyo-dataset}, 2022.

\bibitem[Ceylan et~al.(2023)Ceylan, Huang, and Mitra]{Pix2Video}
Duygu Ceylan, Chun{-}Hao~Paul Huang, and Niloy~J. Mitra.
\newblock Pix2video: Video editing using image diffusion.
\newblock In \emph{Int. Conf. Comput. Vis.}, pages 23149--23160, 2023.

\bibitem[Chai et~al.(2023)Chai, Guo, Wang, and Lu]{StableVideo}
Wenhao Chai, Xun Guo, Gaoang Wang, and Yan Lu.
\newblock Stablevideo: Text-driven consistency-aware diffusion video editing.
\newblock In \emph{Int. Conf. Comput. Vis.}, pages 22983--22993, 2023.

\bibitem[Chan et~al.(2022)Chan, Durand, and Isola]{Line_drawing}
Caroline Chan, Fr{\'e}do Durand, and Phillip Isola.
\newblock Learning to generate line drawings that convey geometry and semantics.
\newblock In \emph{IEEE Conf. Comput. Vis. Pattern Recog.}, pages 7915--7925, 2022.

\bibitem[Chefer et~al.(2024)Chefer, Zada, Paiss, Ephrat, Tov, Rubinstein, Wolf, Dekel, Michaeli, and Mosseri]{Still_Moving}
Hila Chefer, Shiran Zada, Roni Paiss, Ariel Ephrat, Omer Tov, Michael Rubinstein, Lior Wolf, Tali Dekel, Tomer Michaeli, and Inbar Mosseri.
\newblock Still-moving: Customized video generation without customized video data.
\newblock \emph{ACM Trans. Graph.}, 43\penalty0 (6):\penalty0 244:1--244:11, 2024.

\bibitem[Chen et~al.(2023{\natexlab{a}})Chen, Xia, He, Zhang, Cun, Yang, Xing, Liu, Chen, Wang, Weng, and Shan]{VideoCrafter1}
Haoxin Chen, Menghan Xia, Yingqing He, Yong Zhang, Xiaodong Cun, Shaoshu Yang, Jinbo Xing, Yaofang Liu, Qifeng Chen, Xintao Wang, Chao Weng, and Ying Shan.
\newblock Videocrafter1: Open diffusion models for high-quality video generation.
\newblock \emph{CoRR}, abs/2310.19512, 2023{\natexlab{a}}.

\bibitem[Chen et~al.(2024{\natexlab{a}})Chen, Zhang, Cun, Xia, Wang, Weng, and Shan]{VideoCrafter2}
Haoxin Chen, Yong Zhang, Xiaodong Cun, Menghan Xia, Xintao Wang, Chao Weng, and Ying Shan.
\newblock Videocrafter2: Overcoming data limitations for high-quality video diffusion models.
\newblock In \emph{IEEE Conf. Comput. Vis. Pattern Recog.}, pages 7310--7320, 2024{\natexlab{a}}.

\bibitem[Chen et~al.(2024{\natexlab{b}})Chen, Wu, Luo, Xie, Paul, Luo, Zhao, and Li]{PIXART_delta}
Junsong Chen, Yue Wu, Simian Luo, Enze Xie, Sayak Paul, Ping Luo, Hang Zhao, and Zhenguo Li.
\newblock Pixart-{\(\delta\)}: Fast and controllable image generation with latent consistency models.
\newblock \emph{CoRR}, abs/2401.05252, 2024{\natexlab{b}}.

\bibitem[Chen et~al.(2024{\natexlab{c}})Chen, Yu, Ge, Yao, Xie, Wang, Kwok, Luo, Lu, and Li]{PixArt}
Junsong Chen, Jincheng Yu, Chongjian Ge, Lewei Yao, Enze Xie, Zhongdao Wang, James~T. Kwok, Ping Luo, Huchuan Lu, and Zhenguo Li.
\newblock Pixart-{\(\alpha\)}: Fast training of diffusion transformer for photorealistic text-to-image synthesis.
\newblock In \emph{Int. Conf. Learn. Represent.}, 2024{\natexlab{c}}.

\bibitem[Chen et~al.(2020)Chen, Su, Gao, Xia, and Fu]{DeepFaceDrawing}
Shu{-}Yu Chen, Wanchao Su, Lin Gao, Shihong Xia, and Hongbo Fu.
\newblock Deepfacedrawing: deep generation of face images from sketches.
\newblock \emph{ACM Trans. Graph.}, 39\penalty0 (4):\penalty0 72, 2020.

\bibitem[Chen et~al.(2021)Chen, Liu, Lai, Rosin, Li, Fu, and Gao]{DeepFaceEditing}
Shu{-}Yu Chen, Feng{-}Lin Liu, Yu{-}Kun Lai, Paul~L. Rosin, Chunpeng Li, Hongbo Fu, and Lin Gao.
\newblock Deepfaceediting: deep face generation and editing with disentangled geometry and appearance control.
\newblock \emph{ACM Trans. Graph.}, 40\penalty0 (4):\penalty0 90:1--90:15, 2021.

\bibitem[Chen et~al.(2023{\natexlab{b}})Chen, Wu, Xie, Wu, Li, Xia, Xiao, and Lin]{Control-A-Video}
Weifeng Chen, Jie Wu, Pan Xie, Hefeng Wu, Jiashi Li, Xin Xia, Xuefeng Xiao, and Liang Lin.
\newblock Control-a-video: Controllable text-to-video generation with diffusion models, 2023{\natexlab{b}}.

\bibitem[Chen et~al.(2023{\natexlab{c}})Chen, Wang, Zhang, Zhuang, Ma, Yu, Wang, Lin, Qiao, and Liu]{Seine}
Xinyuan Chen, Yaohui Wang, Lingjun Zhang, Shaobin Zhuang, Xin Ma, Jiashuo Yu, Yali Wang, Dahua Lin, Yu Qiao, and Ziwei Liu.
\newblock Seine: Short-to-long video diffusion model for generative transition and prediction.
\newblock In \emph{Int. Conf. Learn. Represent.}, 2023{\natexlab{c}}.

\bibitem[Cheng et~al.(2024)Cheng, Xiao, and He]{InsV2V}
Jiaxin Cheng, Tianjun Xiao, and Tong He.
\newblock Consistent video-to-video transfer using synthetic dataset.
\newblock In \emph{Int. Conf. Learn. Represent.}, 2024.

\bibitem[Chu et~al.(2023)Chu, Lin, and Chen]{Video_ControlNet}
Ernie Chu, Shuo{-}Yen Lin, and Jun{-}Cheng Chen.
\newblock Video controlnet: Towards temporally consistent synthetic-to-real video translation using conditional image diffusion models.
\newblock \emph{CoRR}, abs/2305.19193, 2023.

\bibitem[Cohen et~al.(2024)Cohen, Kulikov, Kleiner, Huberman-Spiegelglas, and Michaeli]{slicedit}
Nathaniel Cohen, Vladimir Kulikov, Matan Kleiner, Inbar Huberman-Spiegelglas, and Tomer Michaeli.
\newblock Slicedit: Zero-shot video editing with text-to-image diffusion models using spatio-temporal slices.
\newblock \emph{arXiv preprint arXiv:2405.12211}, 2024.

\bibitem[Geyer et~al.(2024)Geyer, Bar{-}Tal, Bagon, and Dekel]{TokenFlow}
Michal Geyer, Omer Bar{-}Tal, Shai Bagon, and Tali Dekel.
\newblock Tokenflow: Consistent diffusion features for consistent video editing.
\newblock In \emph{Int. Conf. Learn. Represent.}, 2024.

\bibitem[Goodfellow et~al.(2014)Goodfellow, Pouget{-}Abadie, Mirza, Xu, Warde{-}Farley, Ozair, Courville, and Bengio]{GAN}
Ian~J. Goodfellow, Jean Pouget{-}Abadie, Mehdi Mirza, Bing Xu, David Warde{-}Farley, Sherjil Ozair, Aaron~C. Courville, and Yoshua Bengio.
\newblock Generative adversarial networks.
\newblock \emph{CoRR}, abs/1406.2661, 2014.

\bibitem[Guo et~al.(2024{\natexlab{a}})Guo, Yang, Rao, Agrawala, Lin, and Dai]{SparseCtrl}
Yuwei Guo, Ceyuan Yang, Anyi Rao, Maneesh Agrawala, Dahua Lin, and Bo Dai.
\newblock Sparsectrl: Adding sparse controls to text-to-video diffusion models.
\newblock In \emph{Eur. Conf. Comput. Vis.}, pages 330--348, 2024{\natexlab{a}}.

\bibitem[Guo et~al.(2024{\natexlab{b}})Guo, Yang, Rao, Liang, Wang, Qiao, Agrawala, Lin, and Dai]{AnimateDiff}
Yuwei Guo, Ceyuan Yang, Anyi Rao, Zhengyang Liang, Yaohui Wang, Yu Qiao, Maneesh Agrawala, Dahua Lin, and Bo Dai.
\newblock Animatediff: Animate your personalized text-to-image diffusion models without specific tuning.
\newblock In \emph{Int. Conf. Learn. Represent.}, 2024{\natexlab{b}}.

\bibitem[He et~al.(2024)He, Xu, Guo, Wetzstein, Dai, Li, and Yang]{CameraCtrl}
Hao He, Yinghao Xu, Yuwei Guo, Gordon Wetzstein, Bo Dai, Hongsheng Li, and Ceyuan Yang.
\newblock Cameractrl: Enabling camera control for text-to-video generation.
\newblock \emph{CoRR}, abs/2404.02101, 2024.

\bibitem[He et~al.(2022)He, Yang, Zhang, Shan, and Chen]{LVDM}
Yingqing He, Tianyu Yang, Yong Zhang, Ying Shan, and Qifeng Chen.
\newblock Latent video diffusion models for high-fidelity video generation with arbitrary lengths.
\newblock \emph{CoRR}, abs/2211.13221, 2022.

\bibitem[Ho et~al.(2022)Ho, Salimans, Gritsenko, Chan, Norouzi, and Fleet]{VDM}
Jonathan Ho, Tim Salimans, Alexey Gritsenko, William Chan, Mohammad Norouzi, and David~J Fleet.
\newblock Video diffusion models.
\newblock \emph{Adv. Neural Inform. Process. Syst.}, 35:\penalty0 8633--8646, 2022.

\bibitem[Hong et~al.(2022)Hong, Ding, Zheng, Liu, and Tang]{cogvideo}
Wenyi Hong, Ming Ding, Wendi Zheng, Xinghan Liu, and Jie Tang.
\newblock Cogvideo: Large-scale pretraining for text-to-video generation via transformers.
\newblock \emph{arXiv preprint arXiv:2205.15868}, 2022.

\bibitem[Huang et~al.(2024)Huang, Zhang, and Liao]{LVCD}
Zhitong Huang, Mohan Zhang, and Jing Liao.
\newblock {LVCD:} reference-based lineart video colorization with diffusion models.
\newblock \emph{CoRR}, abs/2409.12960, 2024.

\bibitem[Jamriska et~al.(2019)Jamriska, Sochorov{\'{a}}, Texler, Luk{\'{a}}c, Fiser, Lu, Shechtman, and S{\'{y}}kora]{Stylize_Video_example}
Ondrej Jamriska, S{\'{a}}rka Sochorov{\'{a}}, Ondrej Texler, Michal Luk{\'{a}}c, Jakub Fiser, Jingwan Lu, Eli Shechtman, and Daniel S{\'{y}}kora.
\newblock Stylizing video by example.
\newblock \emph{ACM Trans. Graph.}, 38\penalty0 (4):\penalty0 107:1--107:11, 2019.

\bibitem[Jiang et~al.(2024{\natexlab{a}})Jiang, Zheng, Li, Yang, Wang, and Li]{control_diffusion_survey}
Rui Jiang, Guang-Cong Zheng, Teng Li, Tian-Rui Yang, Jing-Dong Wang, and Xi Li.
\newblock A survey of multimodal controllable diffusion models.
\newblock \emph{Journal of Computer Science and Technology}, 39\penalty0 (3):\penalty0 509--541, 2024{\natexlab{a}}.

\bibitem[Jiang et~al.(2024{\natexlab{b}})Jiang, Wu, Yang, Si, Lin, Qiao, Loy, and Liu]{VideoBooth}
Yuming Jiang, Tianxing Wu, Shuai Yang, Chenyang Si, Dahua Lin, Yu Qiao, Chen~Change Loy, and Ziwei Liu.
\newblock Videobooth: Diffusion-based video generation with image prompts.
\newblock In \emph{IEEE Conf. Comput. Vis. Pattern Recog.}, pages 6689--6700, 2024{\natexlab{b}}.

\bibitem[Kasten et~al.(2021)Kasten, Ofri, Wang, and Dekel]{Layered_Video_Editing}
Yoni Kasten, Dolev Ofri, Oliver Wang, and Tali Dekel.
\newblock Layered neural atlases for consistent video editing.
\newblock \emph{ACM Trans. Graph.}, 40\penalty0 (6):\penalty0 210:1--210:12, 2021.

\bibitem[Khachatryan et~al.(2023)Khachatryan, Movsisyan, Tadevosyan, Henschel, Wang, Navasardyan, and Shi]{Text2Video-Zero}
Levon Khachatryan, Andranik Movsisyan, Vahram Tadevosyan, Roberto Henschel, Zhangyang Wang, Shant Navasardyan, and Humphrey Shi.
\newblock Text2video-zero: Text-to-image diffusion models are zero-shot video generators.
\newblock In \emph{Int. Conf. Comput. Vis.}, pages 15908--15918, 2023.

\bibitem[Ku et~al.(2024)Ku, Wei, Ren, Yang, and Chen]{AnyV2V}
Max Ku, Cong Wei, Weiming Ren, Harry Yang, and Wenhu Chen.
\newblock Anyv2v: A tuning-free framework for any video-to-video editing tasks.
\newblock \emph{arXiv preprint arXiv:2403.14468}, 2024.

\bibitem[Lab and etc.(2024)]{Open-Sora-Plan}
PKU-Yuan Lab and Tuzhan~AI etc.
\newblock Open-sora-plan, 2024.

\bibitem[Li et~al.(2024)Li, Wang, Zhang, Wang, Yuan, Xie, Zou, and Shan]{Image_Conductor}
Yaowei Li, Xintao Wang, Zhaoyang Zhang, Zhouxia Wang, Ziyang Yuan, Liangbin Xie, Yuexian Zou, and Ying Shan.
\newblock Image conductor: Precision control for interactive video synthesis.
\newblock \emph{CoRR}, abs/2406.15339, 2024.

\bibitem[Li et~al.(2023)Li, Zhu, Han, Hou, Guo, and Cheng]{AMT}
Zhen Li, Zuo-Liang Zhu, Ling-Hao Han, Qibin Hou, Chun-Le Guo, and Ming-Ming Cheng.
\newblock Amt: All-pairs multi-field transforms for efficient frame interpolation.
\newblock In \emph{IEEE Conf. Comput. Vis. Pattern Recog.}, 2023.

\bibitem[Lin et~al.(2024)Lin, Liu, Li, and Yang]{zero_SNR}
Shanchuan Lin, Bingchen Liu, Jiashi Li, and Xiao Yang.
\newblock Common diffusion noise schedules and sample steps are flawed.
\newblock In \emph{{IEEE/CVF} Winter Conference on Applications of Computer Vision, {WACV}}, pages 5392--5399, 2024.

\bibitem[Liu et~al.(2022)Liu, Chen, Lai, Li, Jiang, Fu, and Gao]{DFVD}
Feng{-}Lin Liu, Shu{-}Yu Chen, Yu{-}Kun Lai, Chunpeng Li, Yue{-}Ren Jiang, Hongbo Fu, and Lin Gao.
\newblock Deepfacevideoediting: sketch-based deep editing of face videos.
\newblock \emph{ACM Trans. Graph.}, 41\penalty0 (4):\penalty0 167:1--167:16, 2022.

\bibitem[Mou et~al.(2024{\natexlab{a}})Mou, Cao, Wang, Zhang, Shan, and Zhang]{ReVideo}
Chong Mou, Mingdeng Cao, Xintao Wang, Zhaoyang Zhang, Ying Shan, and Jian Zhang.
\newblock Revideo: Remake a video with motion and content control.
\newblock \emph{CoRR}, abs/2405.13865, 2024{\natexlab{a}}.

\bibitem[Mou et~al.(2024{\natexlab{b}})Mou, Wang, Xie, Wu, Zhang, Qi, and Shan]{T2I-Adapter}
Chong Mou, Xintao Wang, Liangbin Xie, Yanze Wu, Jian Zhang, Zhongang Qi, and Ying Shan.
\newblock T2i-adapter: Learning adapters to dig out more controllable ability for text-to-image diffusion models.
\newblock In \emph{AAAI}, pages 4296--4304, 2024{\natexlab{b}}.

\bibitem[Nan et~al.(2024)Nan, Xie, Zhou, Fan, Yang, Chen, Li, Yang, and Tai]{OpenVid_1M}
Kepan Nan, Rui Xie, Penghao Zhou, Tiehan Fan, Zhenheng Yang, Zhijie Chen, Xiang Li, Jian Yang, and Ying Tai.
\newblock Openvid-1m: {A} large-scale high-quality dataset for text-to-video generation.
\newblock \emph{CoRR}, abs/2407.02371, 2024.

\bibitem[OpenAI(2024)]{sora}
OpenAI.
\newblock Sora overview:https://openai.com/index/sora/, 2024.

\bibitem[Ouyang et~al.(2024{\natexlab{a}})Ouyang, Wang, Xiao, Bai, Zhang, Zheng, Zhou, Chen, and Chen]{CoDeF}
Hao Ouyang, Qiuyu Wang, Yuxi Xiao, Qingyan Bai, Juntao Zhang, Kecheng Zheng, Xiaowei Zhou, Qifeng Chen, and Qifeng Chen.
\newblock Codef: Content deformation fields for temporally consistent video processing.
\newblock In \emph{IEEE Conf. Comput. Vis. Pattern Recog.}, pages 8089--8099, 2024{\natexlab{a}}.

\bibitem[Ouyang et~al.(2024{\natexlab{b}})Ouyang, Dong, Yang, Si, and Pan]{I2VEdit}
Wenqi Ouyang, Yi Dong, Lei Yang, Jianlou Si, and Xingang Pan.
\newblock I2vedit: First-frame-guided video editing via image-to-video diffusion models.
\newblock \emph{CoRR}, abs/2405.16537, 2024{\natexlab{b}}.

\bibitem[Parmar et~al.(2024)Parmar, Park, Narasimhan, and Zhu]{One_Step_Translation}
Gaurav Parmar, Taesung Park, Srinivasa Narasimhan, and Jun{-}Yan Zhu.
\newblock One-step image translation with text-to-image models.
\newblock \emph{CoRR}, abs/2403.12036, 2024.

\bibitem[Pavlakos et~al.(2019)Pavlakos, Choutas, Ghorbani, Bolkart, Osman, Tzionas, and Black]{SMPL-X:2019}
Georgios Pavlakos, Vasileios Choutas, Nima Ghorbani, Timo Bolkart, Ahmed A.~A. Osman, Dimitrios Tzionas, and Michael~J. Black.
\newblock Expressive body capture: 3d hands, face, and body from a single image.
\newblock In \emph{IEEE Conf. Comput. Vis. Pattern Recog.}, 2019.

\bibitem[Peng et~al.(2024)Peng, Wang, Zhang, Li, Yang, and Jia]{ControlNeXt}
Bohao Peng, Jian Wang, Yuechen Zhang, Wenbo Li, Ming{-}Chang Yang, and Jiaya Jia.
\newblock Controlnext: Powerful and efficient control for image and video generation.
\newblock \emph{CoRR}, abs/2408.06070, 2024.

\bibitem[Radford et~al.(2021)Radford, Kim, Hallacy, Ramesh, Goh, Agarwal, Sastry, Askell, Mishkin, Clark, Krueger, and Sutskever]{CLIP}
Alec Radford, Jong~Wook Kim, Chris Hallacy, Aditya Ramesh, Gabriel Goh, Sandhini Agarwal, Girish Sastry, Amanda Askell, Pamela Mishkin, Jack Clark, Gretchen Krueger, and Ilya Sutskever.
\newblock Learning transferable visual models from natural language supervision.
\newblock In \emph{ICML}, pages 8748--8763, 2021.

\bibitem[Richardson et~al.(2021)Richardson, Alaluf, Patashnik, Nitzan, Azar, Shapiro, and Cohen{-}Or]{pSp}
Elad Richardson, Yuval Alaluf, Or Patashnik, Yotam Nitzan, Yaniv Azar, Stav Shapiro, and Daniel Cohen{-}Or.
\newblock Encoding in style: {A} stylegan encoder for image-to-image translation.
\newblock In \emph{IEEE Conf. Comput. Vis. Pattern Recog.}, pages 2287--2296, 2021.

\bibitem[Rombach et~al.(2022)Rombach, Blattmann, Lorenz, Esser, and Ommer]{StableDiffusion}
Robin Rombach, Andreas Blattmann, Dominik Lorenz, Patrick Esser, and Bj{\"{o}}rn Ommer.
\newblock High-resolution image synthesis with latent diffusion models.
\newblock In \emph{IEEE Conf. Comput. Vis. Pattern Recog.}, pages 10674--10685, 2022.

\bibitem[Ruder et~al.(2018)Ruder, Dosovitskiy, and Brox]{Style_Transfer_video}
Manuel Ruder, Alexey Dosovitskiy, and Thomas Brox.
\newblock Artistic style transfer for videos and spherical images.
\newblock \emph{Int. J. Comput. Vis.}, 126\penalty0 (11):\penalty0 1199--1219, 2018.

\bibitem[Saharia et~al.(2022)Saharia, Chan, Saxena, Li, Whang, Denton, Ghasemipour, Lopes, Ayan, Salimans, Ho, Fleet, and Norouzi]{DeepFloyd}
Chitwan Saharia, William Chan, Saurabh Saxena, Lala Li, Jay Whang, Emily~L. Denton, Seyed Kamyar~Seyed Ghasemipour, Raphael~Gontijo Lopes, Burcu~Karagol Ayan, Tim Salimans, Jonathan Ho, David~J. Fleet, and Mohammad Norouzi.
\newblock Photorealistic text-to-image diffusion models with deep language understanding.
\newblock In \emph{Adv. Neural Inform. Process. Syst.}, 2022.

\bibitem[Salimans and Ho(2022)]{v_prediction}
Tim Salimans and Jonathan Ho.
\newblock Progressive distillation for fast sampling of diffusion models.
\newblock In \emph{Int. Conf. Learn. Represent.}, 2022.

\bibitem[Schuhmann et~al.(2022)Schuhmann, Beaumont, Vencu, Gordon, Wightman, Cherti, Coombes, Katta, Mullis, Wortsman, Schramowski, Kundurthy, Crowson, Schmidt, Kaczmarczyk, and Jitsev]{LAION_5B}
Christoph Schuhmann, Romain Beaumont, Richard Vencu, Cade Gordon, Ross Wightman, Mehdi Cherti, Theo Coombes, Aarush Katta, Clayton Mullis, Mitchell Wortsman, Patrick Schramowski, Srivatsa Kundurthy, Katherine Crowson, Ludwig Schmidt, Robert Kaczmarczyk, and Jenia Jitsev.
\newblock {LAION-5B:} an open large-scale dataset for training next generation image-text models.
\newblock In \emph{Adv. Neural Inform. Process. Syst.}, 2022.

\bibitem[Shi et~al.(2024)Shi, Huang, Wang, Bian, Li, Zhang, Zhang, Cheung, See, Qin, Dai, and Li]{Motion-I2V}
Xiaoyu Shi, Zhaoyang Huang, Fu{-}Yun Wang, Weikang Bian, Dasong Li, Yi Zhang, Manyuan Zhang, Ka~Chun Cheung, Simon See, Hongwei Qin, Jifeng Dai, and Hongsheng Li.
\newblock Motion-i2v: Consistent and controllable image-to-video generation with explicit motion modeling.
\newblock In \emph{ACM SIGGRAPH}, page 111. {ACM}, 2024.

\bibitem[Singer et~al.(2024)Singer, Zohar, Kirstain, Sheynin, Polyak, Parikh, and Taigman]{EVE}
Uriel Singer, Amit Zohar, Yuval Kirstain, Shelly Sheynin, Adam Polyak, Devi Parikh, and Yaniv Taigman.
\newblock Video editing via factorized diffusion distillation.
\newblock \emph{CoRR}, abs/2403.09334, 2024.

\bibitem[Song et~al.(2021)Song, Meng, and Ermon]{DDIM}
Jiaming Song, Chenlin Meng, and Stefano Ermon.
\newblock Denoising diffusion implicit models.
\newblock In \emph{Int. Conf. Learn. Represent.}, 2021.

\bibitem[Sun et~al.(2024)Sun, Wu, and Gao]{3D_survey}
Jia{-}Mu Sun, Tong Wu, and Lin Gao.
\newblock Recent advances in implicit representation-based 3d shape generation.
\newblock \emph{Vis. Intell.}, 2\penalty0 (1), 2024.

\bibitem[Team and Laboratory(2024)]{Vchitect}
Vchitect Team and Shanghai Artificial~Intelligence Laboratory.
\newblock Vchitect-2.0: Parallel transformer for scaling up video diffusion models, 2024.

\bibitem[Tianpeng et~al.(2024)Tianpeng, Yanxiang, Xinzhe, Yancheng, and Zhiyuan]{Sketch_jig_dataset}
Zheng Tianpeng, Chen Yanxiang, Wen Xinzhe, Li Yancheng, and Wang Zhiyuan.
\newblock Research on diffusion model generated video datasets and detection benchmarks.
\newblock \emph{Journal of Image and Graphics}, pages 1--13, 2024.

\bibitem[Tzaban et~al.(2022)Tzaban, Mokady, Gal, Bermano, and Cohen{-}Or]{Stitch_it_in_Time}
Rotem Tzaban, Ron Mokady, Rinon Gal, Amit Bermano, and Daniel Cohen{-}Or.
\newblock Stitch it in time: Gan-based facial editing of real videos.
\newblock In \emph{{SIGGRAPH} Asia Conference Papers}, pages 29:1--29:9. {ACM}, 2022.

\bibitem[Wang et~al.(2024{\natexlab{a}})Wang, Pan, Peng, Liu, Xu, Miao, Zhan, Tomizuka, Pollefeys, and Wang]{NeRF_survey}
Guangming Wang, Lei Pan, Songyou Peng, Shaohui Liu, Chenfeng Xu, Yanzi Miao, Wei Zhan, Masayoshi Tomizuka, Marc Pollefeys, and Hesheng Wang.
\newblock Nerf in robotics: {A} survey.
\newblock \emph{CoRR}, abs/2405.01333, 2024{\natexlab{a}}.

\bibitem[Wang et~al.(2023)Wang, Yuan, Chen, Zhang, Wang, and Zhang]{ModelScope}
Jiuniu Wang, Hangjie Yuan, Dayou Chen, Yingya Zhang, Xiang Wang, and Shiwei Zhang.
\newblock Modelscope text-to-video technical report.
\newblock \emph{CoRR}, abs/2308.06571, 2023.

\bibitem[Wang et~al.(2018)Wang, Liu, Zhu, Tao, Kautz, and Catanzaro]{pix2pixHD}
Ting{-}Chun Wang, Ming{-}Yu Liu, Jun{-}Yan Zhu, Andrew Tao, Jan Kautz, and Bryan Catanzaro.
\newblock High-resolution image synthesis and semantic manipulation with conditional gans.
\newblock In \emph{IEEE Conf. Comput. Vis. Pattern Recog.}, pages 8798--8807, 2018.

\bibitem[Wang et~al.(2024{\natexlab{b}})Wang, Yuan, Wang, Li, Chen, Xia, Luo, and Shan]{MotionCtrl}
Zhouxia Wang, Ziyang Yuan, Xintao Wang, Yaowei Li, Tianshui Chen, Menghan Xia, Ping Luo, and Ying Shan.
\newblock Motionctrl: {A} unified and flexible motion controller for video generation.
\newblock In \emph{ACM SIGGRAPH}, page 114, 2024{\natexlab{b}}.

\bibitem[Wu et~al.(2024{\natexlab{a}})Wu, Li, Zeng, Zhang, Zhou, Li, Tong, and Chen]{MotionBooth}
Jianzong Wu, Xiangtai Li, Yanhong Zeng, Jiangning Zhang, Qianyu Zhou, Yining Li, Yunhai Tong, and Kai Chen.
\newblock Motionbooth: Motion-aware customized text-to-video generation.
\newblock \emph{CoRR}, abs/2406.17758, 2024{\natexlab{a}}.

\bibitem[Wu et~al.(2024{\natexlab{b}})Wu, Zhang, Wang, Zhou, Zheng, Qi, Shan, and Li]{CustomCrafter}
Tao Wu, Yong Zhang, Xintao Wang, Xianpan Zhou, Guangcong Zheng, Zhongang Qi, Ying Shan, and Xi Li.
\newblock Customcrafter: Customized video generation with preserving motion and concept composition abilities.
\newblock \emph{CoRR}, abs/2408.13239, 2024{\natexlab{b}}.

\bibitem[Xie and Tu(2015)]{HED}
Saining Xie and Zhuowen Tu.
\newblock Holistically-nested edge detection.
\newblock In \emph{Int. Conf. Comput. Vis.}, pages 1395--1403, 2015.

\bibitem[Xing et~al.(2024)Xing, Liu, Xia, Zhang, Wang, Shan, and Wong]{ToonCrafter}
Jinbo Xing, Hanyuan Liu, Menghan Xia, Yong Zhang, Xintao Wang, Ying Shan, and Tien{-}Tsin Wong.
\newblock Tooncrafter: Generative cartoon interpolation.
\newblock \emph{CoRR}, abs/2405.17933, 2024.

\bibitem[Yang et~al.(2023)Yang, Zhou, Liu, and Loy]{Rerender_A_Video}
Shuai Yang, Yifan Zhou, Ziwei Liu, and Chen~Change Loy.
\newblock Rerender {A} video: Zero-shot text-guided video-to-video translation.
\newblock In \emph{{SIGGRAPH} Asia Conference Papers}, pages 95:1--95:11. {ACM}, 2023.

\bibitem[Yang et~al.(2024{\natexlab{a}})Yang, Hou, Huang, Ma, Wan, Zhang, Chen, and Liao]{Direct-a-Video}
Shiyuan Yang, Liang Hou, Haibin Huang, Chongyang Ma, Pengfei Wan, Di Zhang, Xiaodong Chen, and Jing Liao.
\newblock Direct-a-video: Customized video generation with user-directed camera movement and object motion.
\newblock In \emph{ACM SIGGRAPH}, page 113, 2024{\natexlab{a}}.

\bibitem[Yang et~al.(2024{\natexlab{b}})Yang, Teng, Zheng, Ding, Huang, Xu, Yang, Hong, Zhang, Feng, et~al.]{cogvideox}
Zhuoyi Yang, Jiayan Teng, Wendi Zheng, Ming Ding, Shiyu Huang, Jiazheng Xu, Yuanming Yang, Wenyi Hong, Xiaohan Zhang, Guanyu Feng, et~al.
\newblock Cogvideox: Text-to-video diffusion models with an expert transformer.
\newblock \emph{arXiv preprint arXiv:2408.06072}, 2024{\natexlab{b}}.

\bibitem[Yatim et~al.(2024)Yatim, Fridman, Bar-Tal, Kasten, and Dekel]{space_time}
Danah Yatim, Rafail Fridman, Omer Bar-Tal, Yoni Kasten, and Tali Dekel.
\newblock Space-time diffusion features for zero-shot text-driven motion transfer.
\newblock In \emph{IEEE Conf. Comput. Vis. Pattern Recog.}, pages 8466--8476, 2024.

\bibitem[Yuan et~al.(2024{\natexlab{a}})Yuan, Yang, Xiaojuan, and Xiaoping]{Sketch_jig}
Chen Yuan, Zhao Yang, Zhang Xiaojuan, and Liu Xiaoping.
\newblock Sketch colorization with finite color space prior.
\newblock \emph{Journal of Image and Graphics}, 29\penalty0 (4):\penalty0 978--988, 2024{\natexlab{a}}.

\bibitem[Yuan et~al.(2024{\natexlab{b}})Yuan, Yan, Saito, and Fujishiro]{DiffMat}
Liang Yuan, Dingkun Yan, Suguru Saito, and Issei Fujishiro.
\newblock Diffmat: Latent diffusion models for image-guided material generation.
\newblock \emph{Visual Informatics}, 8\penalty0 (1):\penalty0 6--14, 2024{\natexlab{b}}.

\bibitem[Zhang et~al.(2023{\natexlab{a}})Zhang, Rao, and Agrawala]{ControlNet}
Lvmin Zhang, Anyi Rao, and Maneesh Agrawala.
\newblock Adding conditional control to text-to-image diffusion models.
\newblock In \emph{Int. Conf. Comput. Vis.}, pages 3813--3824, 2023{\natexlab{a}}.

\bibitem[Zhang et~al.(2018)Zhang, Isola, Efros, Shechtman, and Wang]{LPIPS}
Richard Zhang, Phillip Isola, Alexei~A. Efros, Eli Shechtman, and Oliver Wang.
\newblock The unreasonable effectiveness of deep features as a perceptual metric.
\newblock In \emph{IEEE Conf. Comput. Vis. Pattern Recog.}, pages 586--595, 2018.

\bibitem[Zhang et~al.(2023{\natexlab{b}})Zhang, Tang, Huang, Huang, Ma, Dong, and Xu]{MotionCrafter}
Yuxin Zhang, Fan Tang, Nisha Huang, Haibin Huang, Chongyang Ma, Weiming Dong, and Changsheng Xu.
\newblock Motioncrafter: One-shot motion customization of diffusion models.
\newblock \emph{CoRR}, abs/2312.05288, 2023{\natexlab{b}}.

\bibitem[Zhang et~al.(2024{\natexlab{a}})Zhang, Wei, Jiang, Zhang, Zuo, and Tian]{ControlVideo}
Yabo Zhang, Yuxiang Wei, Dongsheng Jiang, Xiaopeng Zhang, Wangmeng Zuo, and Qi Tian.
\newblock Controlvideo: Training-free controllable text-to-video generation.
\newblock In \emph{Int. Conf. Learn. Represent.}, 2024{\natexlab{a}}.

\bibitem[Zhang et~al.(2024{\natexlab{b}})Zhang, Liao, Li, Qin, and Wang]{Tora}
Zhenghao Zhang, Junchao Liao, Menghao Li, Long Qin, and Weizhi Wang.
\newblock Tora: Trajectory-oriented diffusion transformer for video generation.
\newblock \emph{CoRR}, abs/2407.21705, 2024{\natexlab{b}}.

\bibitem[Zhang et~al.(2024{\natexlab{c}})Zhang, Wu, Wang, Luo, Zhang, Zhao, Vajda, Metaxas, and Yu]{AVID}
Zhixing Zhang, Bichen Wu, Xiaoyan Wang, Yaqiao Luo, Luxin Zhang, Yinan Zhao, Peter Vajda, Dimitris~N. Metaxas, and Licheng Yu.
\newblock {AVID:} any-length video inpainting with diffusion model.
\newblock In \emph{IEEE Conf. Comput. Vis. Pattern Recog.}, pages 7162--7172, 2024{\natexlab{c}}.

\bibitem[Zhao et~al.(2023)Zhao, Chen, Chen, Bao, Hao, Yuan, and Wong]{Unicontrol}
Shihao Zhao, Dongdong Chen, Yen{-}Chun Chen, Jianmin Bao, Shaozhe Hao, Lu Yuan, and Kwan{-}Yee~K. Wong.
\newblock Uni-controlnet: All-in-one control to text-to-image diffusion models.
\newblock In \emph{Adv. Neural Inform. Process. Syst.}, 2023.

\bibitem[Zheng et~al.(2024)Zheng, Peng, Yang, Shen, Li, Liu, Zhou, Li, and You]{opensora}
Zangwei Zheng, Xiangyu Peng, Tianji Yang, Chenhui Shen, Shenggui Li, Hongxin Liu, Yukun Zhou, Tianyi Li, and Yang You.
\newblock Open-sora: Democratizing efficient video production for all, 2024.

\bibitem[Zhu et~al.(2017)Zhu, Park, Isola, and Efros]{CycleGAN}
Jun{-}Yan Zhu, Taesung Park, Phillip Isola, and Alexei~A. Efros.
\newblock Unpaired image-to-image translation using cycle-consistent adversarial networks.
\newblock In \emph{Int. Conf. Comput. Vis.}, pages 2242--2251, 2017.

\bibitem[Zhu et~al.(2024)Zhu, Wang, Kong, and Wang]{3DGS_survey}
Siting Zhu, Guangming Wang, Dezhi Kong, and Hesheng Wang.
\newblock 3d gaussian splatting in robotics: {A} survey.
\newblock \emph{CoRR}, abs/2410.12262, 2024.

\end{thebibliography}


\begin{thebibliography}{19}
\providecommand{\natexlab}[1]{#1}
\providecommand{\url}[1]{\texttt{#1}}
\expandafter\ifx\csname urlstyle\endcsname\relax
  \providecommand{\doi}[1]{doi: #1}\else
  \providecommand{\doi}{doi: \begingroup \urlstyle{rm}\Url}\fi

\bibitem[Chen et~al.(2024)Chen, Wu, Luo, Xie, Paul, Luo, Zhao, and Li]{PIXART_delta}
Junsong Chen, Yue Wu, Simian Luo, Enze Xie, Sayak Paul, Ping Luo, Hang Zhao, and Zhenguo Li.
\newblock Pixart-{\(\delta\)}: Fast and controllable image generation with latent consistency models.
\newblock \emph{CoRR}, abs/2401.05252, 2024.

\bibitem[Chen et~al.(2023)Chen, Wang, Zhang, Zhuang, Ma, Yu, Wang, Lin, Qiao, and Liu]{Seine}
Xinyuan Chen, Yaohui Wang, Lingjun Zhang, Shaobin Zhuang, Xin Ma, Jiashuo Yu, Yali Wang, Dahua Lin, Yu Qiao, and Ziwei Liu.
\newblock Seine: Short-to-long video diffusion model for generative transition and prediction.
\newblock In \emph{Int. Conf. Learn. Represent.}, 2023.

\bibitem[Cheng et~al.(2024)Cheng, Xiao, and He]{InsV2V}
Jiaxin Cheng, Tianjun Xiao, and Tong He.
\newblock Consistent video-to-video transfer using synthetic dataset.
\newblock In \emph{Int. Conf. Learn. Represent.}, 2024.

\bibitem[Geyer et~al.(2024)Geyer, Bar{-}Tal, Bagon, and Dekel]{TokenFlow}
Michal Geyer, Omer Bar{-}Tal, Shai Bagon, and Tali Dekel.
\newblock Tokenflow: Consistent diffusion features for consistent video editing.
\newblock In \emph{Int. Conf. Learn. Represent.}, 2024.

\bibitem[Guo et~al.(2024)Guo, Yang, Rao, Agrawala, Lin, and Dai]{SparseCtrl}
Yuwei Guo, Ceyuan Yang, Anyi Rao, Maneesh Agrawala, Dahua Lin, and Bo Dai.
\newblock Sparsectrl: Adding sparse controls to text-to-video diffusion models.
\newblock In \emph{Eur. Conf. Comput. Vis.}, pages 330--348, 2024.

\bibitem[Hong et~al.(2022)Hong, Ding, Zheng, Liu, and Tang]{cogvideo}
Wenyi Hong, Ming Ding, Wendi Zheng, Xinghan Liu, and Jie Tang.
\newblock Cogvideo: Large-scale pretraining for text-to-video generation via transformers.
\newblock \emph{arXiv preprint arXiv:2205.15868}, 2022.

\bibitem[Huang et~al.(2022)Huang, Zhang, Heng, Shi, and Zhou]{huang2022rife}
Zhewei Huang, Tianyuan Zhang, Wen Heng, Boxin Shi, and Shuchang Zhou.
\newblock Real-time intermediate flow estimation for video frame interpolation.
\newblock In \emph{Eur. Conf. Comput. Vis.}, 2022.

\bibitem[Ku et~al.(2024)Ku, Wei, Ren, Yang, and Chen]{AnyV2V}
Max Ku, Cong Wei, Weiming Ren, Harry Yang, and Wenhu Chen.
\newblock Anyv2v: A tuning-free framework for any video-to-video editing tasks.
\newblock \emph{arXiv preprint arXiv:2403.14468}, 2024.

\bibitem[Li et~al.(2023)Li, Zhu, Han, Hou, Guo, and Cheng]{AMT}
Zhen Li, Zuo-Liang Zhu, Ling-Hao Han, Qibin Hou, Chun-Le Guo, and Ming-Ming Cheng.
\newblock Amt: All-pairs multi-field transforms for efficient frame interpolation.
\newblock In \emph{IEEE Conf. Comput. Vis. Pattern Recog.}, 2023.

\bibitem[Loshchilov and Hutter(2019)]{AdamW}
Ilya Loshchilov and Frank Hutter.
\newblock Decoupled weight decay regularization.
\newblock In \emph{Int. Conf. Learn. Represent.}, 2019.

\bibitem[Nan et~al.(2024)Nan, Xie, Zhou, Fan, Yang, Chen, Li, Yang, and Tai]{OpenVid_1M}
Kepan Nan, Rui Xie, Penghao Zhou, Tiehan Fan, Zhenheng Yang, Zhijie Chen, Xiang Li, Jian Yang, and Ying Tai.
\newblock Openvid-1m: {A} large-scale high-quality dataset for text-to-video generation.
\newblock \emph{CoRR}, abs/2407.02371, 2024.

\bibitem[Ouyang et~al.(2024)Ouyang, Dong, Yang, Si, and Pan]{I2VEdit}
Wenqi Ouyang, Yi Dong, Lei Yang, Jianlou Si, and Xingang Pan.
\newblock I2vedit: First-frame-guided video editing via image-to-video diffusion models.
\newblock \emph{CoRR}, abs/2405.16537, 2024.

\bibitem[Schuhmann et~al.(2022)Schuhmann, Beaumont, Vencu, Gordon, Wightman, Cherti, Coombes, Katta, Mullis, Wortsman, Schramowski, Kundurthy, Crowson, Schmidt, Kaczmarczyk, and Jitsev]{LAION_5B}
Christoph Schuhmann, Romain Beaumont, Richard Vencu, Cade Gordon, Ross Wightman, Mehdi Cherti, Theo Coombes, Aarush Katta, Clayton Mullis, Mitchell Wortsman, Patrick Schramowski, Srivatsa Kundurthy, Katherine Crowson, Ludwig Schmidt, Robert Kaczmarczyk, and Jenia Jitsev.
\newblock {LAION-5B:} an open large-scale dataset for training next generation image-text models.
\newblock In \emph{Adv. Neural Inform. Process. Syst.}, 2022.

\bibitem[Wang et~al.(2021)Wang, Xie, Dong, and Shan]{Real-ESRGAN}
Xintao Wang, Liangbin Xie, Chao Dong, and Ying Shan.
\newblock Real-esrgan: Training real-world blind super-resolution with pure synthetic data.
\newblock In \emph{ICCVW}, pages 1905--1914, 2021.

\bibitem[Xing et~al.(2024)Xing, Liu, Xia, Zhang, Wang, Shan, and Wong]{ToonCrafter}
Jinbo Xing, Hanyuan Liu, Menghan Xia, Yong Zhang, Xintao Wang, Ying Shan, and Tien{-}Tsin Wong.
\newblock Tooncrafter: Generative cartoon interpolation.
\newblock \emph{CoRR}, abs/2405.17933, 2024.

\bibitem[Yan et~al.(2023{\natexlab{a}})Yan, Zhang, Fan, and Wu]{UCF_YAN_ICCV2023}
Zhiyuan Yan, Yong Zhang, Yanbo Fan, and Baoyuan Wu.
\newblock Ucf: Uncovering common features for generalizable deepfake detection.
\newblock In \emph{Int. Conf. Comput. Vis.}, pages 22412--22423, 2023{\natexlab{a}}.

\bibitem[Yan et~al.(2023{\natexlab{b}})Yan, Zhang, Yuan, Lyu, and Wu]{DeepfakeBench_YAN_NEURIPS2023}
Zhiyuan Yan, Yong Zhang, Xinhang Yuan, Siwei Lyu, and Baoyuan Wu.
\newblock Deepfakebench: A comprehensive benchmark of deepfake detection.
\newblock In \emph{Adv. Neural Inform. Process. Syst.}, pages 4534--4565, 2023{\natexlab{b}}.

\bibitem[Yan et~al.(2024)Yan, Luo, Lyu, Liu, and Wu]{LSDA_YAN_CVPR2024}
Zhiyuan Yan, Yuhao Luo, Siwei Lyu, Qingshan Liu, and Baoyuan Wu.
\newblock Transcending forgery specificity with latent space augmentation for generalizable deepfake detection.
\newblock In \emph{IEEE Conf. Comput. Vis. Pattern Recog.}, 2024.

\bibitem[Zhang et~al.(2023)Zhang, Rao, and Agrawala]{ControlNet}
Lvmin Zhang, Anyi Rao, and Maneesh Agrawala.
\newblock Adding conditional control to text-to-image diffusion models.
\newblock In \emph{Int. Conf. Comput. Vis.}, pages 3813--3824, 2023.

\end{thebibliography}
}

\end{document}

\clearpage
\setcounter{page}{1}
\maketitlesupplementary

In this supplemental material, we provide additional implementation details (Sec. \ref{sec:Implementation_details}), showcase results of sketch-based video generation (Sec. \ref{sec:supple_generation_results}) and editing (Sec. \ref{sec:supple_editing_results}), and the results of consecutive generation and editing (Sec. \ref{sec:supple_synthetic_editing}). 
We also present additional ablation study (Sec. \ref{sec:supple_ablation_fusion}), {memory and time analysis (Sec. \ref{sec:supple_memory_time_efficiency}),} comparison results (Sec.  \ref{sec:supple_compare_editing}){, attention map visualization (Sec. \ref{sec:supple_propagation_illustration}),} and failure cases (Sec. \ref{sec:failure_cases}).
We discuss ethical concerns in Sec. \ref{sec:supple_ethical_discuss} and show the full text prompts of all examples in Sec. \ref{sec:full_text_prompt}. 

\vspace{-3mm}

\section{Implementation Details}
\label{sec:Implementation_details}

\textbf{Sketch-based Generation.}
During training, we use a learning rate of 1e-5 and employ mixed precision (fp16). 
We optimize the sketch-conditioned network using the AdamW optimizer \cite{AdamW}, while fixing the weights of the pretrained CogVideoX-2b \cite{cogvideo} model. %
Similar to ControlNet \cite{ControlNet}, we initialize the weights of the DiT block with those from CogVideoX-2b.
In the hybrid training stage (image and video), each batch consists solely of images or videos, with one or two keyframe sketches conditioned on random time points. 
The training dataset includes 53w video clips and 90w images, with the labeled text prompts from OpenVid \cite{OpenVid_1M} and LAION \cite{LAION_5B}.

During inference, generated videos have a resolution of 720×480 and 8 fps, spanning 6 seconds with 49 frames. 
The classifier-free guidance scale is set to 10.0.
Super-resolution \cite{Real-ESRGAN} and frame interpolation \cite{huang2022rife} can be optionally applied to improve the resolution and frame rate, as in CogVideo.
All the results shown in the paper and supplemental material do not use them. 

\textbf{Sketch-based Editing.}
We initialize the editing network using the weights from the sketch-based generation model. 
During training, the rectangle masks have random position, height, and weight. 
Mask movement strategies include:
1) fixed position, 2) linear interpolation between two random positions, and 3) motion based on the mean optical flow of the pixels within the box. 
The masks are directly multiplied with the input videos at the pixel level to remove the information in the edited regions. 

During inference, the classifier-free guidance scale is set to 20.0.
Users may manually adjust mask positions through interpolation or dynamic movement using optical flow.
The masks are further downsampled and merged to achieve latent fusion. 
The 3D VAE in CogVideoX-2b performs temporally 8×8 spatial and 4× temporal downsampling. So, we also downsample the masks by 8×8 to match the latent code resolution and combine the masks of four frames corresponding to one latent code.
The masks are shown with orange boxes in the original videos in Fig. \ref{fig:supple_editing} and %
\ref{fig:supple_gen_editing}.

\section{Sketch-based Generation Results}
\label{sec:supple_generation_results}
\begin{figure*}[h]
    \centering
    \includegraphics[width=1.0\linewidth]{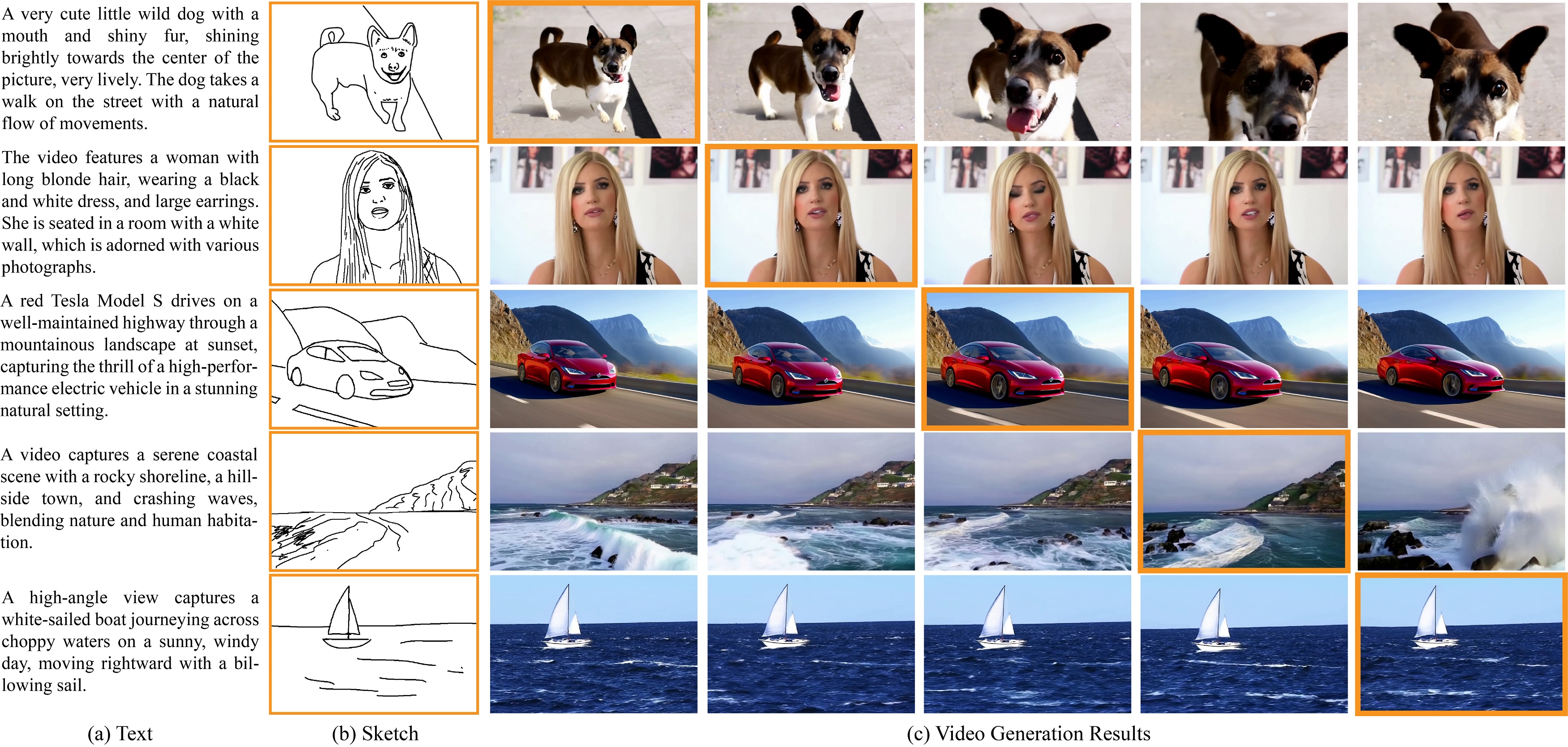}
    \caption{Video generation results using one keyframe sketch. Given the text prompts (a) and sketches (b), our method generates realistic video results (c). The frames corresponding to the given time points are highlighted in orange boxes. 
    }
    \label{fig:supple_one_key}
\end{figure*}

\begin{figure*}[h]
    \centering
    \includegraphics[width=0.9\linewidth]{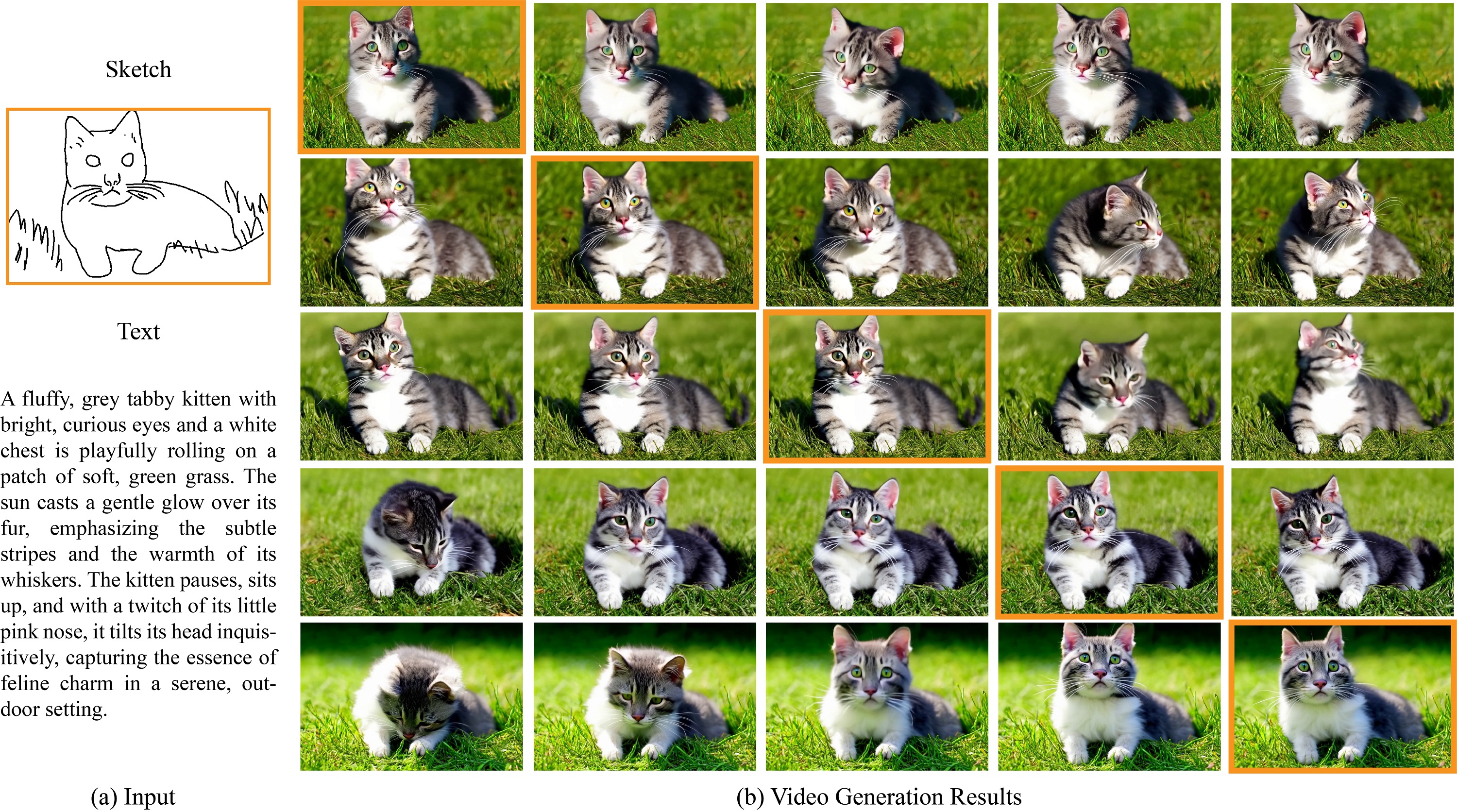}
    \caption{Video generation results using the same sketch and text inputs (a) with different time points. The frames at 0s, 1.5s, 3s, 4.5s, and 6s (from top to bottom) are marked with orange boxes. 
    The results (b) have good faithfulness to the sketch at given time points, while exhibiting diverse motion in other frames. 
    }
    \label{fig:supple_one_key_diverse_timepoint}
\end{figure*}

We show additional sketch-based video generation results to validate the effectiveness of our method. 
All results in this section are generated from \textbf{hand-drawn sketches} rather than synthetic ones. 
Fig. \ref{fig:supple_one_key} shows the video generation results using a single keyframe sketch as input, conditioned on different time points (0s, 1.5s, 3s, 4.5s, and 6s). Our method generates realistic video results across diverse categories. 
Fig. \ref{fig:supple_one_key_diverse_timepoint} shows the results from the same sketch and text prompts, but with varying time points (0s, 1.5s, 3s, 4.5s, and 6s). 
These results exhibit fidelity to the sketch at the specified time point while introducing diverse motion in other frames.

\begin{figure*}[h]
    \centering
    \includegraphics[width=1.0\linewidth]{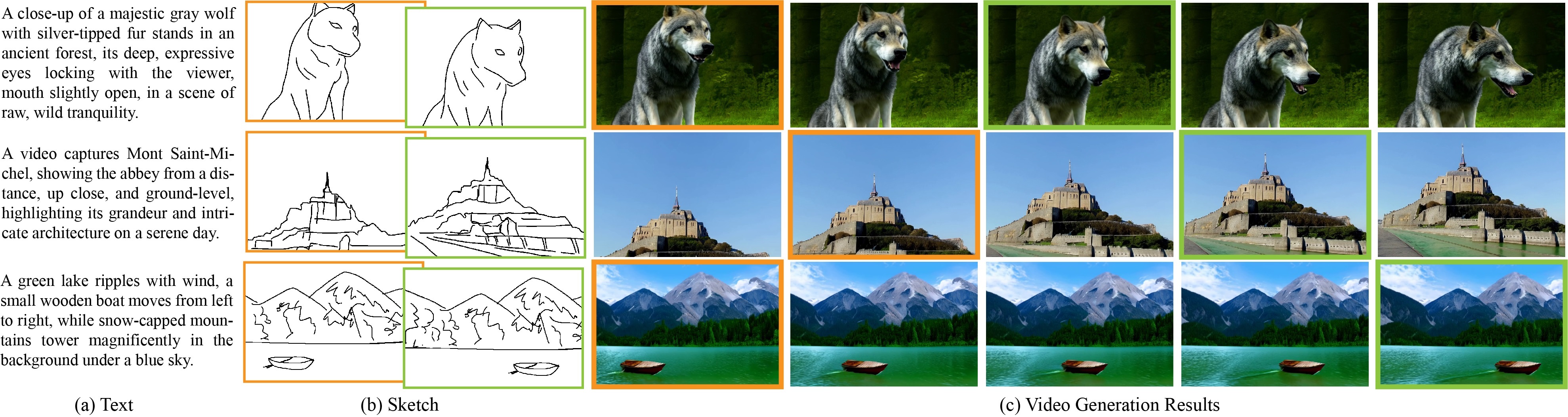}
    \caption{Video generation results using two keyframe sketches. Given the text prompts (a) and two sketches (b), our method generates realistic video results (c). The frames corresponding to the given time points are highlighted in orange and green boxes.
    }
    \label{fig:supple_two_key}
\end{figure*}

\begin{figure*}[h]
    \centering
    \includegraphics[width=0.9\linewidth]{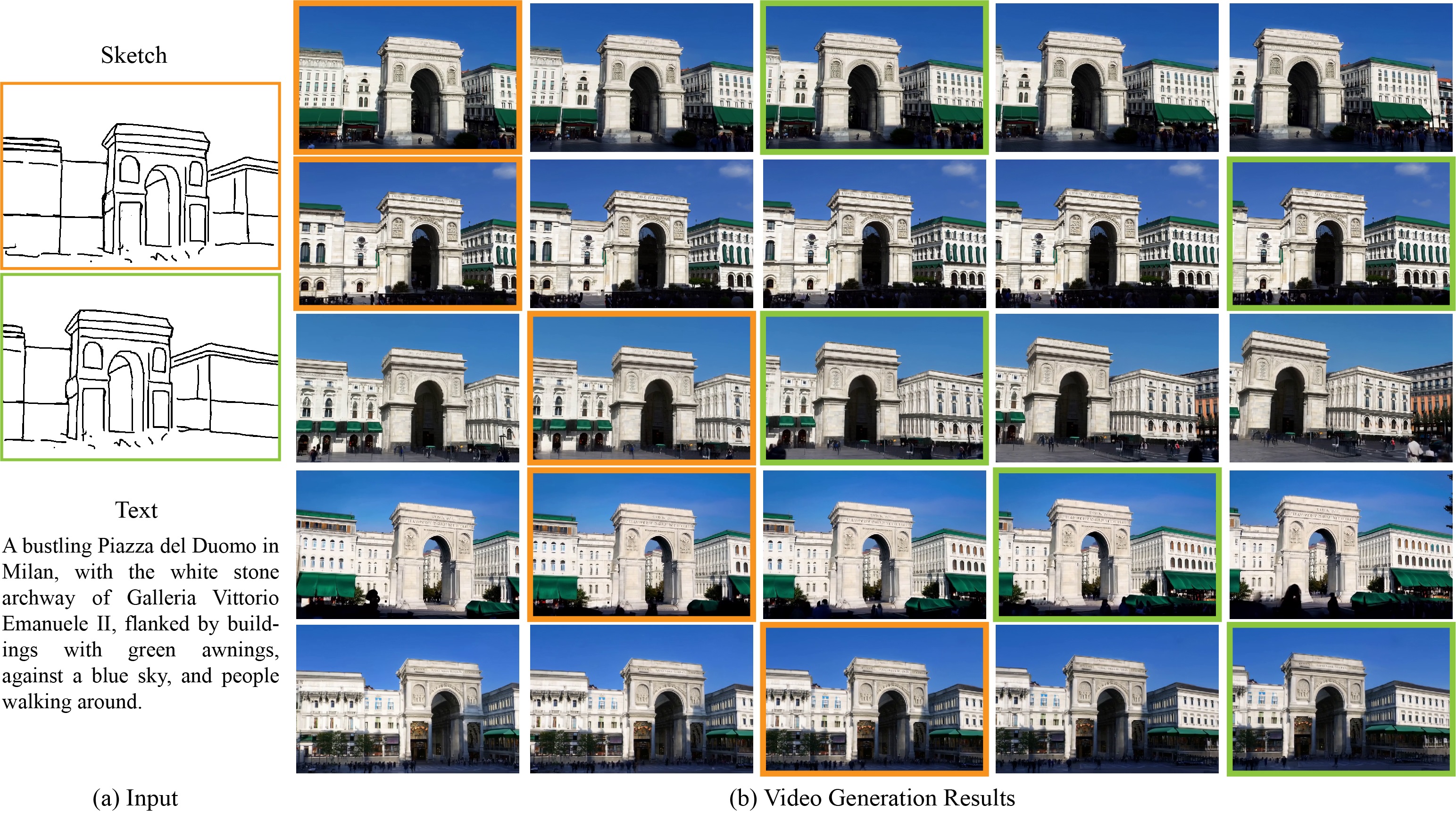}
    \caption{Video generation results (b) using the same two keyframe sketches and text inputs (a) at diverse time points. %
    Each column in (b) represents frames at 0s, 1.5s, 3s, 4.5s, and 6s (from left to right). The frames corresponding to the given time points are highlighted in orange and green boxes. %
    }
    \label{fig:supple_two_key_diverse_timepoint}
\end{figure*}

In Fig. \ref{fig:supple_two_key}, we present additional video generation results from two keyframe sketches. 
Our method generates realistic videos with detailed control of geometry and motion. 
Fig. \ref{fig:supple_two_key_diverse_timepoint} shows the results with the same two keyframe sketches and text prompts, but varying time points. 
It can be seen that our method generates interesting interpolation and extrapolation results. 

\begin{figure*}[h]
    \centering
    \includegraphics[width=1.0\linewidth]{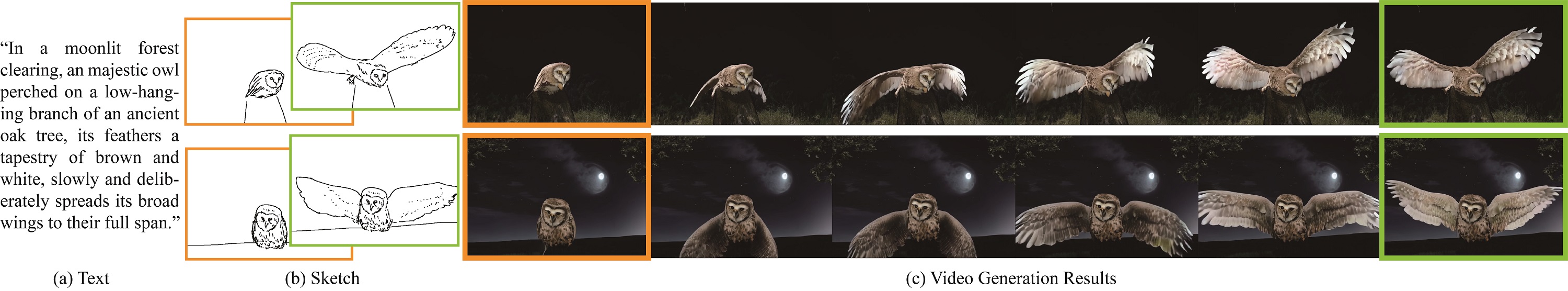}
    \caption{Video generation results using the same text prompts (a) and different keyframe sketches (b). %
    The generated results (c) have a similar appearance while exhibiting different geometry details. 
    The keyframe sketches and corresponding generated frames are highlighted in orange and green box.
    }
    \label{fig:Diverse_sketch}
\end{figure*}

\begin{figure*}[h]
    \centering
    \includegraphics[width=1.0\linewidth]{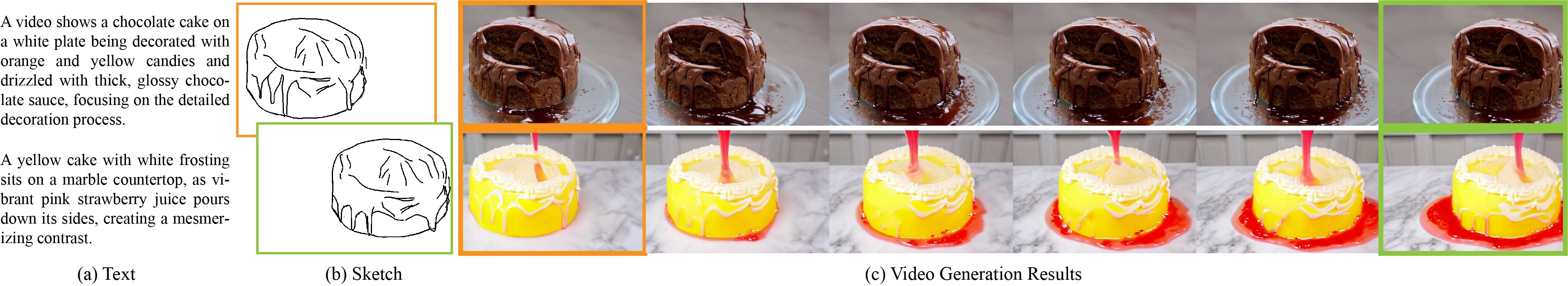}
    \caption{The video generation results using the same two keyframe sketches %
    (b) and different text prompts (a).
    The generated results (c) share a similar structure while exhibiting different appearances, demonstrating the influence of text prompts on visual details. 
    The keyframe sketches and corresponding generated frames are highlighted in the orange and green boxes.
    }
    \label{fig:Diverse_text}
\end{figure*}

Our method uses the sketch to define geometry and the text prompt to specify appearance.
As shown in Fig. \ref{fig:Diverse_sketch}, given the identical text but varying sketches, our method generates results of similar objects but with different geometry components and layouts. 
As shown in Fig. \ref{fig:Diverse_text}, our method generates diverse results from identical sketches with different text prompts, varying the appearance while maintaining the same structure.

\section{Sketch-based Editing Results}
\label{sec:supple_editing_results}

\begin{figure*}[h]
    \centering
    \includegraphics[width=1.0\linewidth]{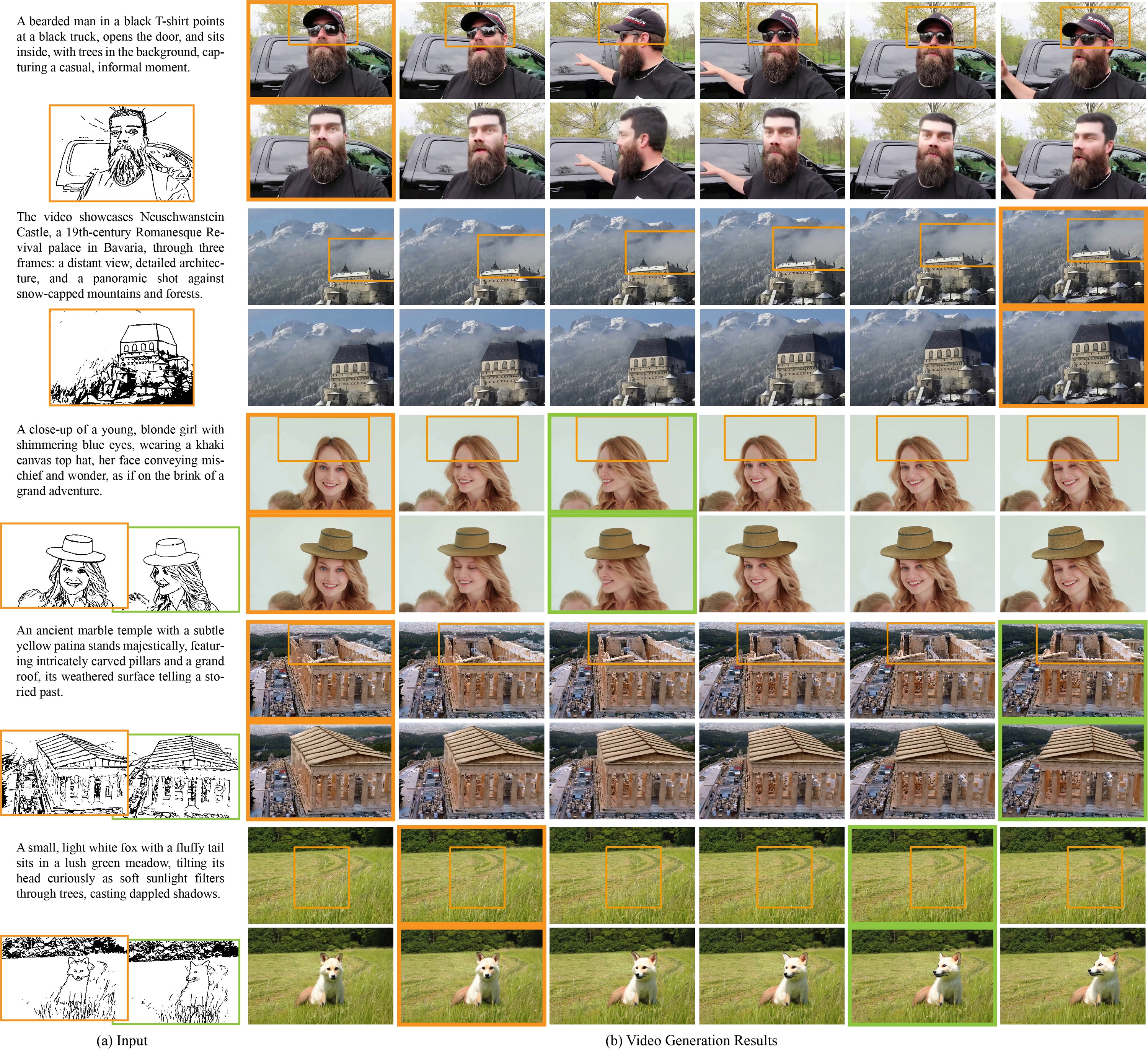}
    \caption{Results of sketch-based video editing. For each example, %
    the input text and drawing sketch(es) %
    are given in (a). In (b), the original video is at the top, while the edited video is at the bottom. 
    The sketches and corresponding frames are highlighted in orange and green boxes.
    }
    \label{fig:supple_editing}
\end{figure*}

We show additional results of sketch-base video editing in Fig. \ref{fig:supple_editing}. 
Our method effectively edits object components within the video, such as modifying a person's face or changing the roof of a castle. 
Our method also supports editing based on two keyframe sketches at different time points, controlling the motion of the newly generated objects (e.g., the transformation of a hat or the posing of a fox).

\section{Consecutive Editing Results}
\label{sec:supple_synthetic_editing}

\begin{figure*}[h]
    \centering
    \includegraphics[width=1.0\linewidth]{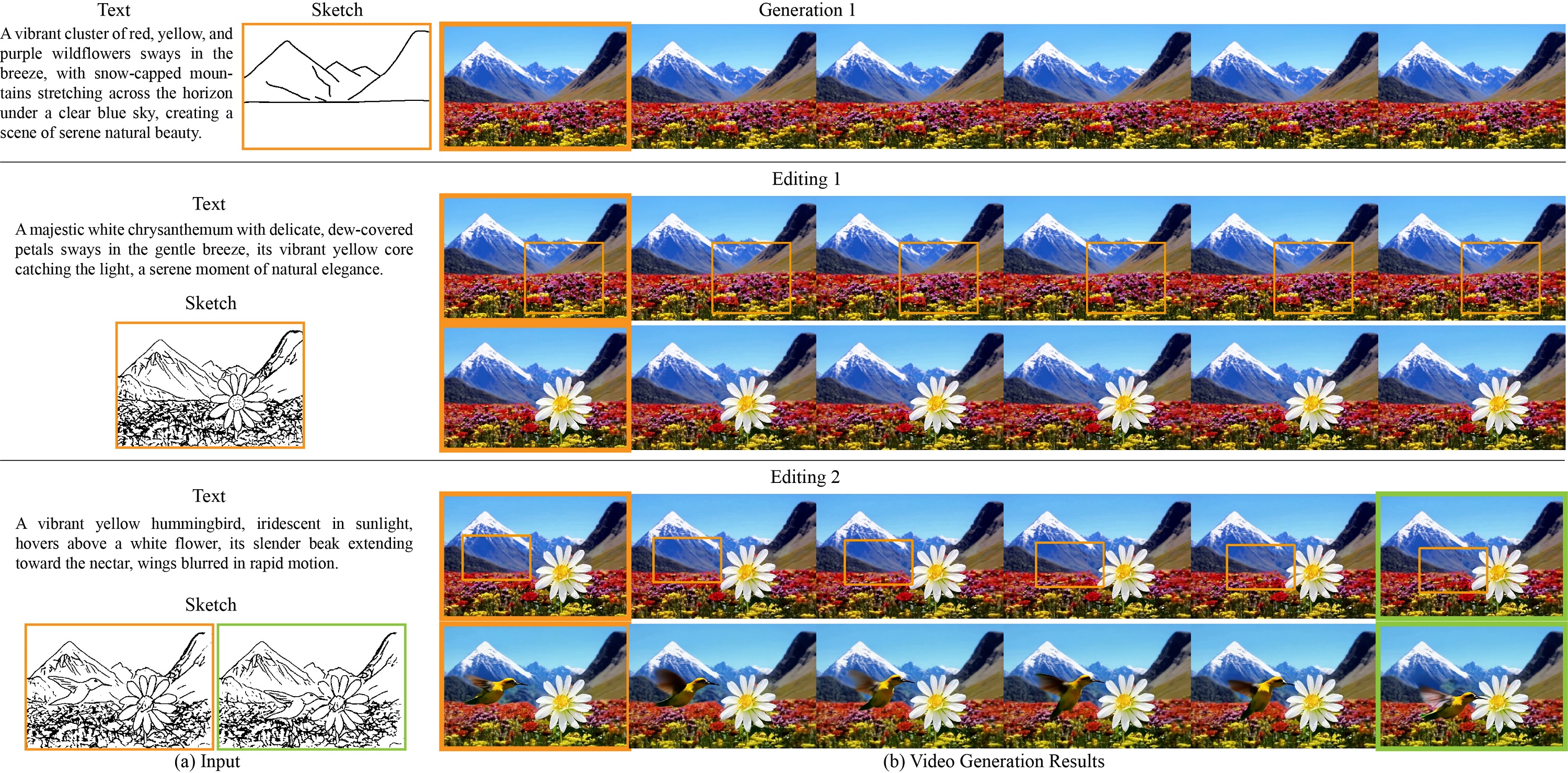}
    \caption{Consecutive video generation and editing results.
    In Step 1 (Generation 1), a video is generated based on an input text and a sketch. 
    In Step 2 (Editing 1), a flower is added to %
    the generated video.
    Step 3 (Editing 2) adds a hummingbird to the edited video. 
    }
    \label{fig:supple_gen_editing}
\end{figure*}

Our method allows for consecutive generation and editing of videos. 
As shown in Fig. \ref{fig:supple_gen_editing}, given an initial sketch and a text prompt, we first generate a realistic video (Generation 1). 
The generated %
video then serves as input for further editing, using a new text prompt and a new sketch %
to modify the content (Editing 1).
Additional edits (Editing 2) can be made iteratively without degrading the quality of the generated video. 
This showcases the detailed control our method offers for video customization.

\section{More Ablation Study}
\label{sec:supple_ablation_fusion}
\begin{figure*}[h]
    \centering
    \includegraphics[width=0.9\linewidth]{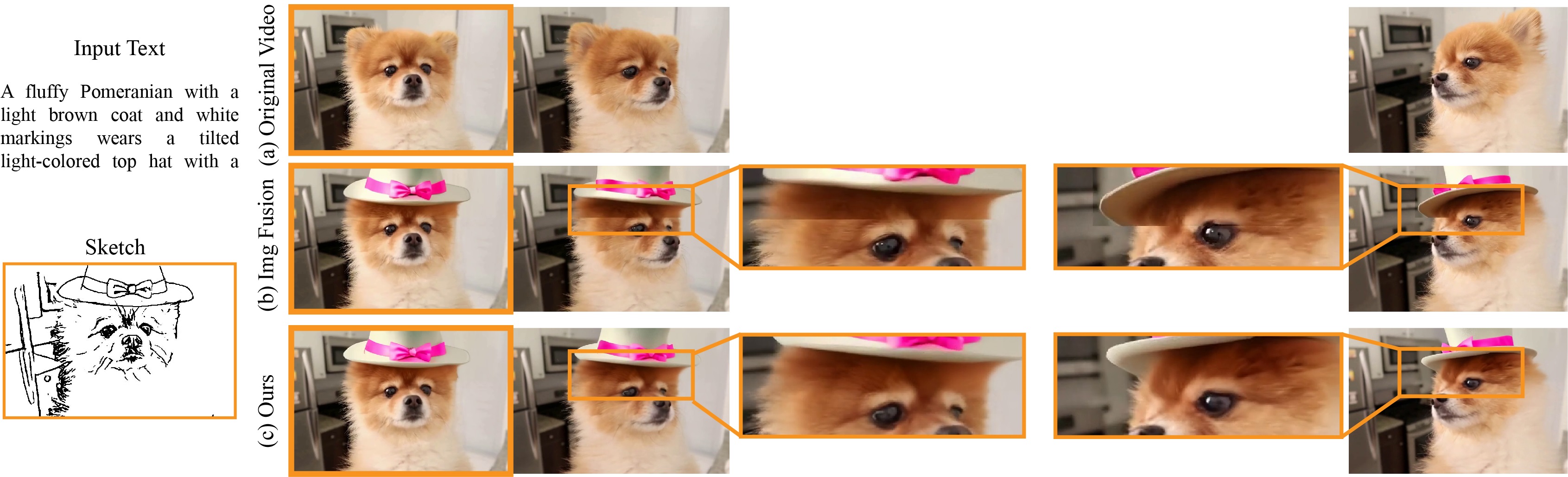}
    \caption{Ablation study of latent fusion. Replacing the latent fusion strategy with direct image space fusion leads to edited results (b) with noticeable seams. 
    Our method removes this artifact and generates a realistic edited video with the new component {added} into the original content.
    }
    \label{fig:supple_ablation_fusion}
\end{figure*}

During the editing process, we employ a latent fusion strategy to seamlessly fuse the newly generated regions with the original video content. 
An alternative approach involves directly fusing the generated video with the input video in pixel level.
As shown in Fig. \ref{fig:supple_ablation_fusion} (b), this baseline approach results in obvious seams along the boundaries of the edited regions. 
In contrast, our method ensures a smooth fusion of the edited and unedited regions by accurately inserting information from the unedited regions during the denoising process. 
And the combination in latent space followed with VAE decoding generates more realistic images than direct pixel level fusion. 

\begin{table}[h]\small %
    \centering
    \vspace{2mm}
    \begin{tabular}{ccccc}
        \hline
        \makebox[0.07\textwidth][c]{Metric} & \makebox[0.07\textwidth][c] {1-5} & \makebox[0.07\textwidth][c] {13-17} & \makebox[0.07\textwidth][c] {26-30} & \makebox[0.07\textwidth][c] {Ours}\\
        \hline
        LPIPS$\downarrow$ & 33.37& 38.57& 50.34& \textbf{32.47}\\
        \hline
    \end{tabular}
    \caption{{Different distributions of sketch control blocks. Our method has the best sketch faithfulness compared with other distributions.}}
    \label{tab:distributed_table}
\end{table}

\begin{table}[h]\footnotesize
    \begin{center}
    \resizebox{\columnwidth}{!}{
    \begin{tabular}{lccc|cccc}
        \toprule
         & \multicolumn{3}{c}{\textit{Training}} & \multicolumn{4}{c}{\textit{Inference}}\\
        \textit{Method} & {Ctrl-10} &{Ctrl-5} & {Ours} & SpCtrl & Ctrl-10 & Ctrl-5 & Ours \\\midrule
        Memory(GB) \quad
        & 70.79 & 55.37 & 56.84 & 11.94 & 21.23 & 19.95 & 20.35 \\
        Time(s) \quad
        & N/A & N/A & N/A & 29 & 52 & 46 & 41 \\
        \hline
    \end{tabular}}
    \caption{{The memory and time efficiency of different methods. SpCtrl refers to SparseCtrl~\cite{SparseCtrl} method. Ctrl-10 and Ctrl-5 refer to the Ctrl-CogVideo baseline with 10 and 5 control blocks, respectively.}}
    \label{tab:memory_time_table}
    \end{center}
\end{table}

\textbf{Distribution of control blocks.} 
We constructed a control branch with blocks at positions 1-5, 13-17, 26-30, and a uniform distribution (ours). 
To evaluate the effectiveness of these placements, we compute LPIPS between the input and extracted sketches from corresponding generated frames.
Due to the restriction of computing resources, we trained 10,000 steps on our hybrid image and video dataset.
As shown in Table \ref{tab:distributed_table}, our method incorporates control signals at multiple feature levels, enabling a more comprehensive analysis and achieving superior performance compared to the %
baselines. Additionally, in the DiT-based video diffusion model (CogVideo-2b in our method), we observe that earlier blocks primarily influence geometric features, while later blocks exhibit reduced controllability in this aspect.

\section{Memory and Time Efficiency}
\label{sec:supple_memory_time_efficiency}
A direct comparison of memory and time with existing methods, such as SparseCtrl, is not meaningful due to different architectures (Unet for SparseCtrl, DiT for ours) and frame counts (16 vs. 49 frames). 
We applied SparseCtrl's concept to CogVideo, but using half of the blocks (15 blocks) as in Pixart-$\delta$~\cite{PIXART_delta} resulted in out-of-memory errors.
In Table \ref{tab:memory_time_table}, we report its memory and time with 10 blocks (Ctrl-10) and 5 blocks (Ctrl-5) below. 
Our sketch control block consumes slightly more memory compared with Ctrl-5 due to additional inter-frame attention.
Our method achieves faster inference speed compared with the Ctrl-5 and Ctrl-10 baselines. %

\begin{table*}\small %
  \centering
  \begin{tabular}{ccccccc}
    \toprule
    \makebox[0.1\textwidth][c]{Metrics} & \makebox[0.1\textwidth][c]{ToonCrafter\cite{ToonCrafter}} & \makebox[0.1\textwidth][c]{Seine\cite{Seine}} & 
    \makebox[0.1\textwidth][c]{AMT\cite{AMT}} & \makebox[0.1\textwidth][c]{SparseCtrl\cite{SparseCtrl}} & \makebox[0.12\textwidth][c]{Ctrl-CogVideo} & \makebox[0.1\textwidth][c]{Ours} \\ 
    \midrule
    {LPIPS $\downarrow$} & 29.09 & 29.54 & 29.17 & 44.85 & 32.23 & \cellcolor{orange}{27.56} \\
    {CLIP $\uparrow$} & 95.75 & 92.76 & 96.12 & 96.48 & 98.04 & \cellcolor{orange}{98.31} \\
    \bottomrule
  \end{tabular}
  \caption{
  {The quantitative results of sketch-based video generation comparison. The LPIPS and CLIP numbers are scaled up 100×, with each cell colored to indicate the \colorbox{orange}{best}. Methods ToonCrafter \cite{ToonCrafter}, Seine \cite{Seine}, and AMT \cite{AMT} %
  interpolate ControlNet-generated images while the other method %
  directly translate sketches into videos.}
  }
  \label{tab:comp_gen}
\end{table*}

\begin{table*}\small %
  \centering
  \begin{tabular}{ccccccc}
    \toprule
    \makebox[0.1\textwidth][c]{Metrics} & \makebox[0.1\textwidth][c]{InsV2V\cite{InsV2V}} & \makebox[0.1\textwidth][c]{AnyV2V\cite{AnyV2V}} & 
    \makebox[0.1\textwidth][c]{I2VEdit\cite{I2VEdit}} & \makebox[0.1\textwidth][c]{TokenFlow\cite{TokenFlow}} & \makebox[0.1\textwidth][c]{Ours} \\ 
    \midrule
    {LPIPS $\downarrow$} & 13.61 & 11.92 & 10.98 & 12.92 & \cellcolor{orange}{9.74} \\
    {CLIP $\uparrow$} & 95.39 & 93.47 & 96.81 & 97.83 & \cellcolor{orange}{98.34} \\
    {PSNR $\uparrow$} & 16.84 & 13.68 & 12.98 & 21.15 & \cellcolor{orange}{36.48} \\
    \bottomrule
  \end{tabular}
  \caption{
  {The quantitative results of sketch-based video editing comparison. The LPIPS and CLIP numbers are scaled up 100×, with each cell colored to indicate the \colorbox{orange}{best}. AnyV2V \cite{AnyV2V} and I2VEdit \cite{I2VEdit} propagate the image editing results into videos. InV2V \cite{InsV2V} and TokenFLow \cite{TokenFlow} utilize text prompts for video editing.}
  }
  \label{tab:comp_gen}
\end{table*}

\section{More Comparison Results}
\label{sec:supple_compare_editing}
We present additional comparison results for both generation and editing. 
For sketch-based generation, Fig. \ref{fig:supple_gen_compare_1frame} shows comparison results of a single keyframe sketch. 
We exclude results from {video interpolation methods~\cite{AMT, Seine, ToonCrafter}} %
since {they require} %
two keyframes.
Our method generates results that are most faithful to the input sketches, such as the wing shape and a notch in the cake. 
Fig. \ref{fig:supple_gen_compare_2frame} shows comparison results for two keyframe sketches.
{We compare with two additional video interpolation methods, including Seine~\cite{Seine} and ToonCrafter~\cite{ToonCrafter}, whose official codes and weights are utilized in our experiments. Since the text-to-image results of two keyframes are not consistent, these methods generate fuzzy details in intermediate frames.}
Our method generates the most realistic results with the best temporal coherence. 

For sketch-based editing, Fig. \ref{fig:supple_gen_compare_editing} shows additional comparison results, {including two additional methods: TokenFlow~\cite{TokenFlow} and I2VEdit~\cite{I2VEdit}}.
Our method generates the most realistic and geometrically consistent results. In contrast, {TokenFlow~\cite{TokenFlow} exhibits minor editing effects due to the complexity of geometry modifications, while InsV2V~\cite{InsV2V} has a similar problem with insufficient geometry control. Furthermore, AnyV2V~\cite{AnyV2V} and I2VEdit~\cite{I2VEdit} totally change the unedited regions and have fuzzy details.}

\textbf{Quantitative Comparison.}
We conducted quantitative experiments, including additional generation methods (i.e., Seine~\cite{Seine} and ToonCrafter~\cite{ToonCrafter}) and editing methods (i.e., TokenFlow~\cite{TokenFlow} and I2VEdit~\cite{I2VEdit}). For generations, Seine and ToonCrafter achieved similar sketch faithfulness to AMT, as they utilize the same beginning and ending frames. However, they exhibit lower temporal consistency due to the smaller number of frames and greater content variation between adjacent frames.
For editing, existing methods significantly change the unedited regions, as reflected in their low reconstruction values in these regions. Our method has the best sketch faithfulness, temporal consistency, and unedited region preservation. 

\section{Sketch Propagation Illustration}
\label{sec:supple_propagation_illustration}
\begin{figure}[h]
    \centering
    \includegraphics[width=1.0\linewidth]{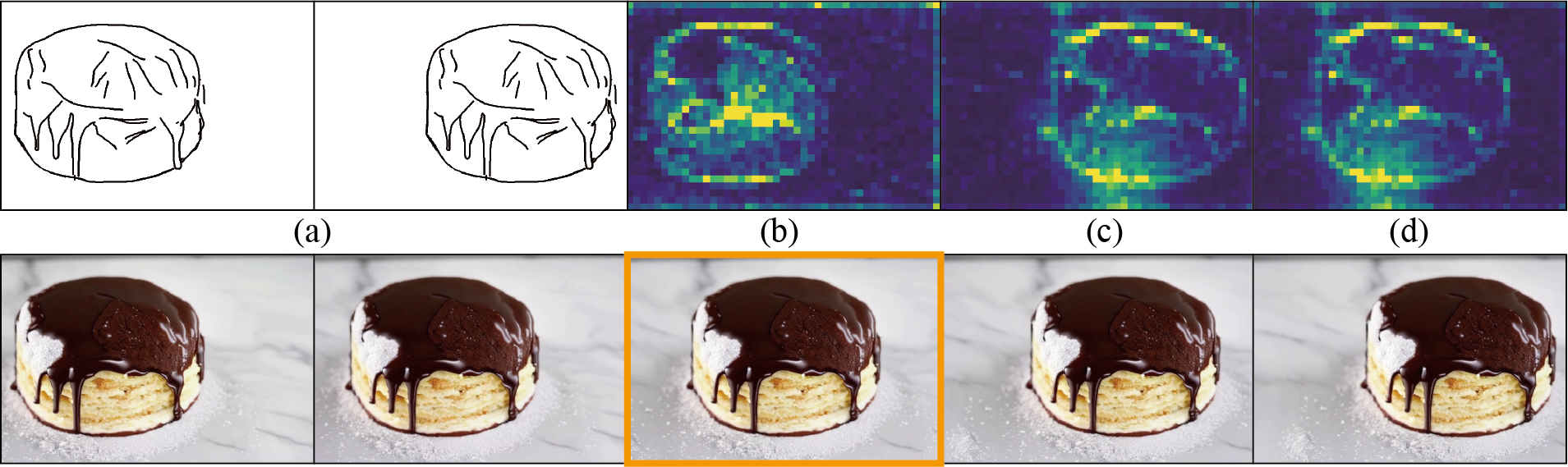}
    \caption{
    {The visualization of inter-frame attention maps. The input sketches (a) and attention maps (b,c,d) are shown in the top row, while the generated video is shown in the bottom row.}
    }
    \label{fig:supple_visual}
\end{figure}

\begin{figure*}[h]
    \centering
    \includegraphics[width=1.0\linewidth]{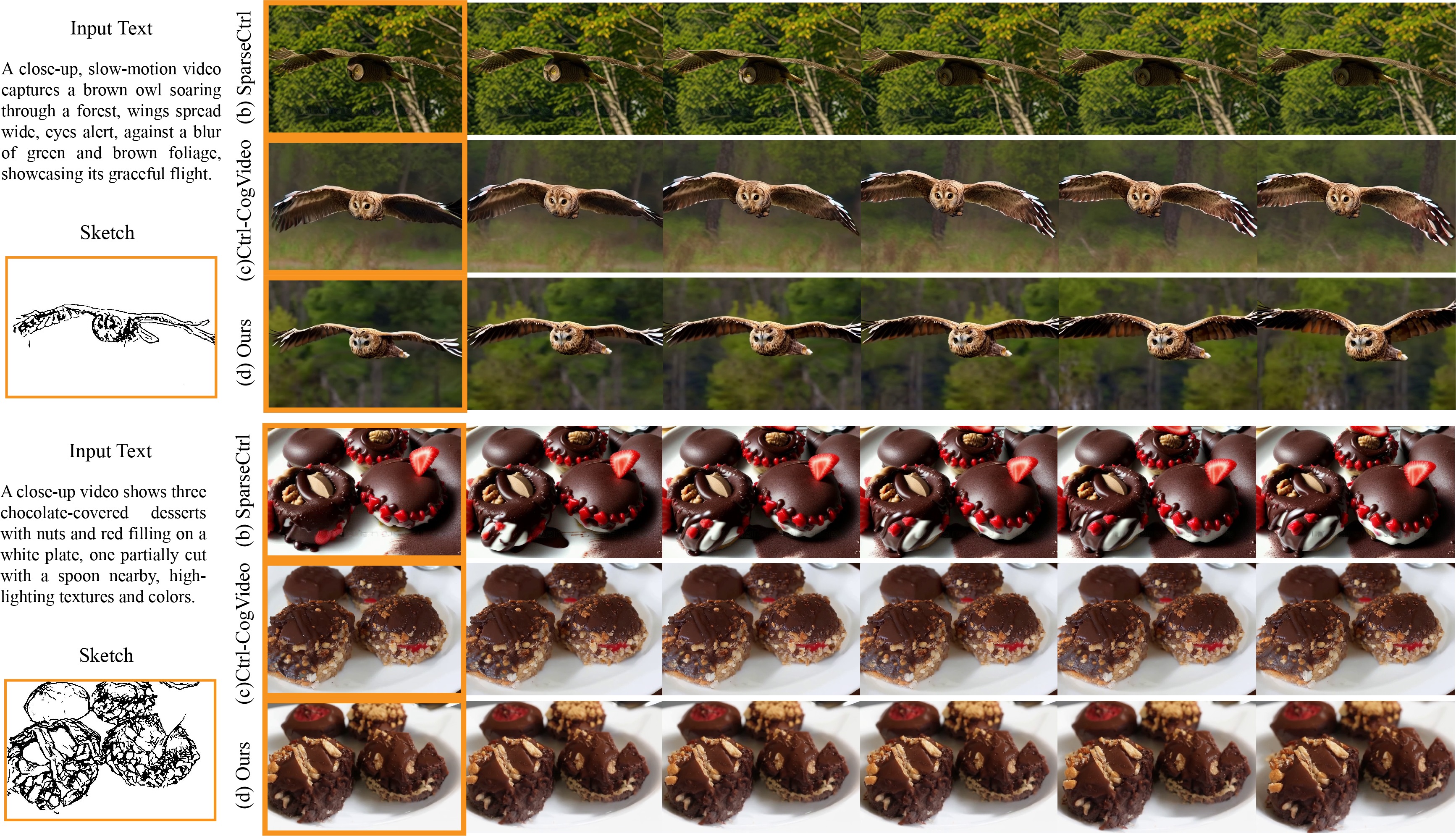}
    \caption{Comparison of sketch-based video generation results for one keyframe sketch input. We compare our method with SparseCtrl \cite{SparseCtrl} and Ctrl-CogVideo (baseline discussion in Sec 4.3 in main paper).
    Our method generates the results that are most faithful to the input sketches. 
    }
    \label{fig:supple_gen_compare_1frame}
\end{figure*}

\begin{figure*}[h]
    \centering
    \includegraphics[width=1.0\linewidth]{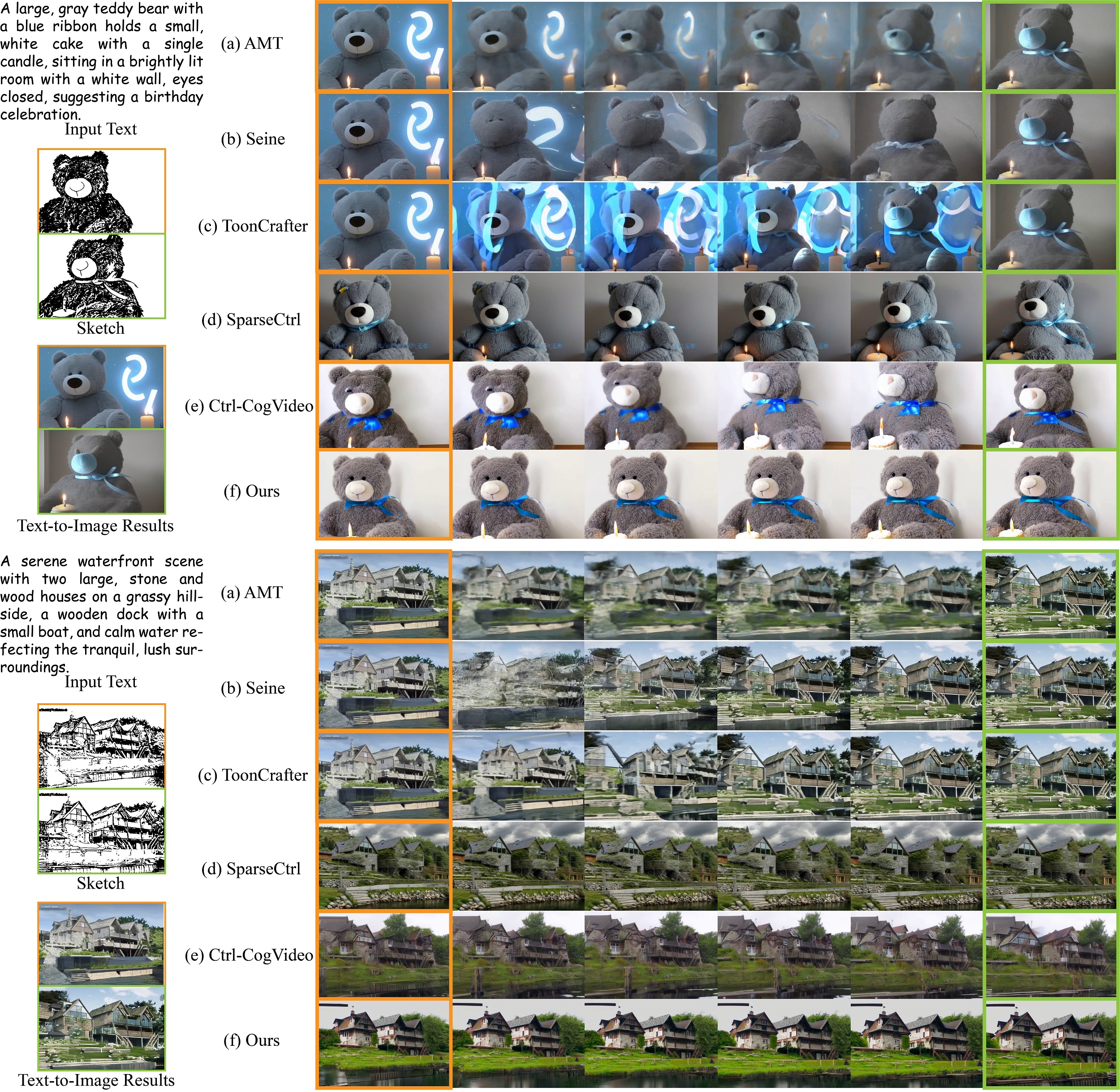}
    \caption{Comparison of sketch-based video generation results for two keyframe sketches. 
    The input text prompts and sketches are shown in the left region, while the results of AMT \cite{AMT}, {Seine \cite{Seine}, ToonCrafter \cite{ToonCrafter},} SparseCtrl \cite{SparseCtrl}, Ctrl-CogVideo (baseline discussed in Sec 4.3 in the main paper), and our method are shown in the right region. 
    Our method achieves the best temporal coherence in the generated video. 
    }
    \label{fig:supple_gen_compare_2frame}
\end{figure*}

\begin{figure*}[h]
    \centering
    \includegraphics[width=1.0\linewidth]{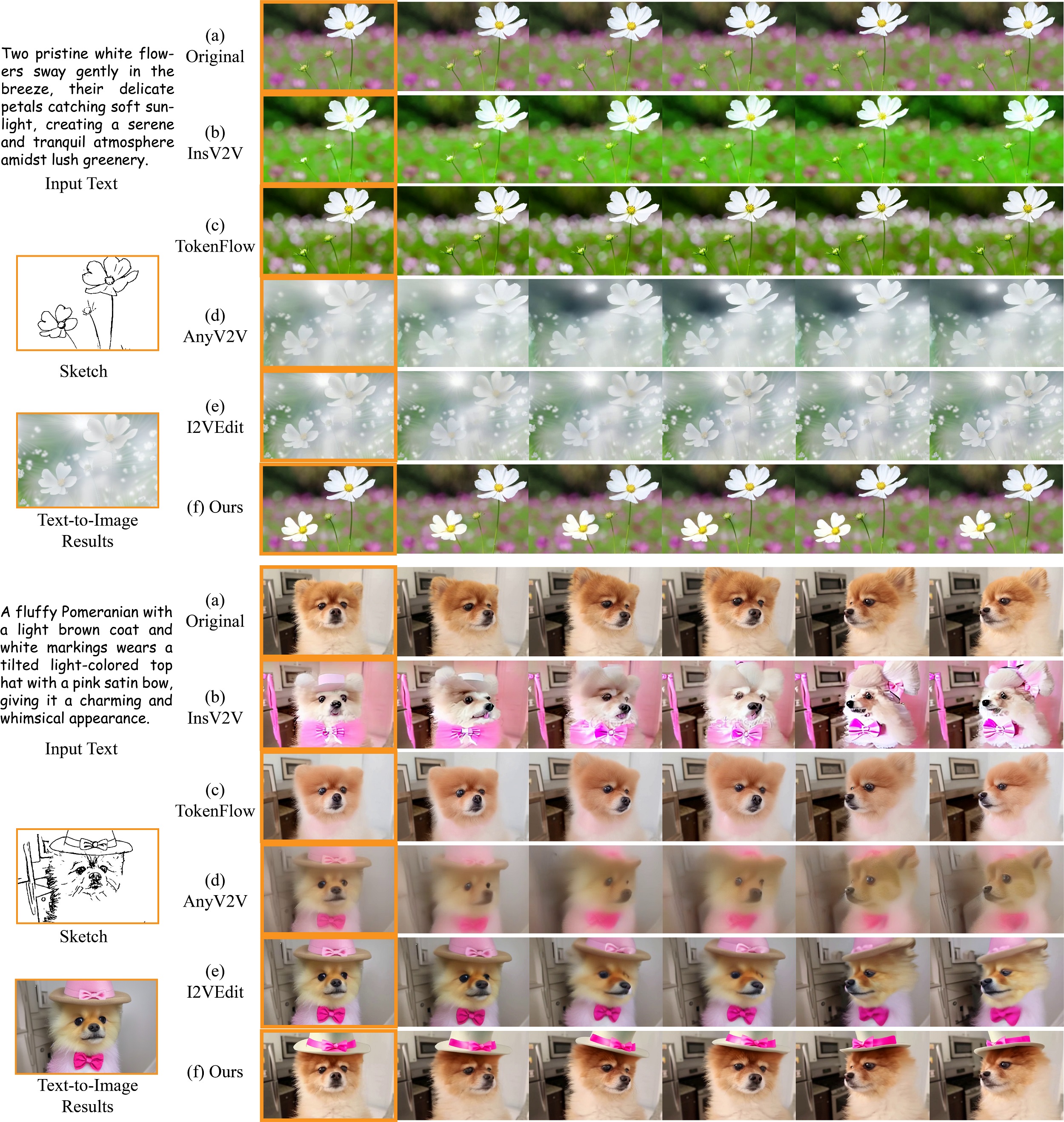}
    \caption{Comparison of video editing results for different methods. 
    The input text prompts and sketches are shown on the left. The edited results from InsV2V \cite{InsV2V}, {TokenFlow \cite{TokenFlow}}, AnyV2V \cite{AnyV2V}, {I2VEdit \cite{I2VEdit}}, and ours are shown on the right. 
    Our method generates the most realistic video editing results. 
    }
    \label{fig:supple_gen_compare_editing}
\end{figure*}

As shown in Fig. \ref{fig:supple_visual}, we visualize the attention maps during the generation of {a frame highlighted in an orange box}, both for input sketches (b, c) and itself (d). 
They are consistent with {the} input sketches and {frame's edge maps}, 
supporting our inter-frame attention mechanism for edge analysis and sketch-based controllability. 

\section{Failure Cases}
\label{sec:failure_cases}

\begin{figure*}[h]
    \centering
    \includegraphics[width=1.0\linewidth]{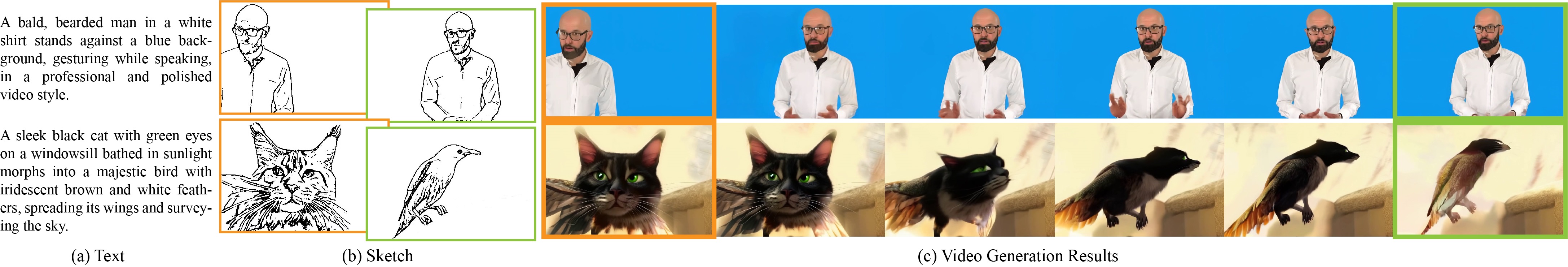}
    \caption{Failure cases of our method. Our method generates fuzzy details in hands and eyeglasses. 
    Performance degrades when the two keyframe sketches (highlighted in orange and green box) %
    depict entirely different objects, resulting in strange transitions between frames.
    }
    \label{fig:supple_failure_cases}
\end{figure*}

Fig. \ref{fig:supple_failure_cases} shows %
failure cases of our method. %
Similar to image generation, our method generates fuzzy details in challenging cases, such as the human hands. 
Additionally, when the two keyframe sketches differ significantly in content, the generated video transitions may appear unnatural or incoherent due to the lack of such a dataset.
3D geometry information and data augmentation can be potentially used to solve these problems. 

\section{Ethical Discussion}
\label{sec:supple_ethical_discuss}

Our method is designed for positive applications such as movie and short video production.
We strongly condemn the abuse of AI video generation technologies.
The methods \cite{DeepfakeBench_YAN_NEURIPS2023, UCF_YAN_ICCV2023, LSDA_YAN_CVPR2024} that detect the fake video may be helpful to avoid the potential misuse of existing video generation approaches. 
Our method can also be used to generate training data for these detection methods.

\section{Full Text Prompt}
\label{sec:full_text_prompt}

For ease of reading, the text prompts shown in the figures are simplified. The full-text prompts of each figure are shown in the following:

\textbf{Main Paper}

Figure 1: 

The camera meticulously frames a close-up of a majestic male lion's face, his mane a regal cascade of tawny hues, eyes a piercing amber gaze that speaks of the wild spirit within. The sunlight bathes his features in a golden glow, highlighting the rugged texture of his fur and the subtle scars that tell tales of past battles. His whiskers are pronounced, sensitive to the faintest breeze, and his mouth is slightly agape, revealing formidable teeth, a silent testament to his power as the king of the savannah.

The video is a drone shot of a serene landscape featuring a deep blue lagoon surrounded by lush green trees and rocky cliffs. The lagoon is nestled between the mountains, creating a picturesque scene. The drone captures the tranquility of the water and the vibrant colors of the surrounding nature. The video is a beautiful representation of the natural beauty of the area, showcasing the lagoon's unique location and the stunning views it offers.

The video captures the breathtaking beauty of a mountainous landscape. The first frame shows a clear blue sky above the mountains, with a few clouds scattered across it. The second frame reveals a deep blue lake nestled between the mountains, its calm waters reflecting the surrounding scenery. The third frame offers a closer view of the mountains, showcasing their rugged terrain and the patches of snow that still cling to their peaks. The overall style of the video is serene and majestic, capturing the natural beauty of the landscape in a way that is both awe-inspiring and tranquil.

A magnificent waterfall cascades down from a series of rocky cliffs. The waterfall occupies most of the area in the picture, which is very magnificent and spectacular. The water flow of the waterfall is very turbulent, with white splashes spanning the entire scene. The waterfall cascades from the top of the cliff all the way to the bottom of the lake.

A fluffy Pomeranian dog wears a light-colored top hat made of cloth adorned with a pink satin bow tied neatly around the base of the hat. The dog has a light brown coat with white markings on its face and chest. The top hat, slightly tilted, gives the Pomeranian a charming and playful appearance, while the pink bow adds a touch of elegance and whimsy. The dog’s fur is well-groomed, showcasing its signature soft, voluminous coat that frames its small face.

Figure 3:

A small, fluffy rabbit with soft, white fur and a pink nose hops gracefully from left to right across a lush, green meadow. The sun is setting, casting a warm, golden glow over the scene. The rabbit pauses momentarily, its ears twitching as it listens to the gentle rustling of the leaves in the nearby trees. It then continues its playful dance, moving side to side with an air of innocence and joy, the grass beneath its paws a soft carpet of nature's embrace.

A sweeping panoramic video captures the grandeur of a majestic city, where towering skyscrapers pierce the clouds, their glass facades shimmering in the sunlight. Nestled at the base of these architectural giants, quaint urban homes with red-tiled roofs and leafy gardens form a charming tapestry around the city's perimeter. In the bottom left corner, a tranquil lake mirrors the sky with its rippling waves of azure water, while fluffy white clouds drift lazily across the expansive blue canvas above, completing the serene yet powerful urban landscape.

The video captures a close-up of a cat's face in three frames. The cat has striking green eyes and a brown and black spotted coat. In the first frame, the cat is looking directly at the camera, its eyes wide and alert. In the second frame, the cat's gaze is slightly averted, and its eyes are more relaxed. In the third frame, the cat is looking away from the camera, its eyes focused on something in the distance. The style of the video is a simple, yet intimate portrait of the cat, with a focus on its expressive eyes and fur pattern.

In a breathtaking cosmic event, a video captures the catastrophic explosion of a light blue planet, its surface fragmenting as rocks are catapulted into the void. The sky behind is a tapestry of stars, while the planet itself becomes a maelstrom of debris, with towering plumes of smoke and cascades of fiery orange flames painting a harrowing picture of celestial destruction. The detritus from the disintegrating planet creates a mesmerizing display against the serene backdrop of space, a stark contrast between the calm universe and the violent upheaval of the once-stable world.

The video is a time-lapse aerial shot of a large, ornate building with a distinctive dome and multiple balconies. The building is surrounded by lush greenery and a well-manicured lawn. In the first frame, the building is bathed in daylight, with the sun shining brightly on its facade. In the second frame, the sun begins to set, casting a warm glow on the building and its surroundings. In the third frame, the sun has set, and the building is illuminated by artificial lights, creating a stark contrast with the darkening sky. The style of the video is realistic, with a focus on the architectural details of the building and the natural beauty of its surroundings.

Figure 4:

Plumes of steam begin to billow from the volcano's crest, hinting at the raw power simmering beneath the surface, a prelude to the impending eruption that promises to reshape the landscape.Rising majestically in the distance, a colossal volcano stands as a silent sentinel, its peak dusted with a delicate blanket of white snow. The volcano's flanks are lush with verdant vegetation, a vibrant green tapestry that contrasts starkly against the earthy tones of its rocky surface. In the foreground of the screen, the lake is rippling with water waves

A man is speaking. He has thick hair and is wearing a black suit with a white shirt and a red tie underneath. The camera focuses on his face and upper body.

The ornamental fish swims from left to right. A mesmerizing ornamental fish glides through the inky depths of the ocean, its vibrant scales shimmering with hues of electric blue, fiery orange, and iridescent green. With a sleek, flattened body that navigates the vast underwater expanse. Its delicate fins, adorned with intricate patterns, flutter and sway like rhythmic poetry in motion, leaving a trail of iridescence in the silent sea.

A green grassland and a small path extending into the distance. The grassland is covered with gravel.

Figure 5:

The video captures a large cruise ship docked at a pier. The ship is painted in white and black, with a distinctive yellow and gold design on its side. The ship is moored to the pier, which is lined with several smaller boats. The sky above is filled with clouds, suggesting an overcast day. The water around the ship is calm, reflecting the ship's grandeur. The pier extends into the water, providing a clear view of the ship's size and scale. The video is a time-lapse, capturing the ship's arrival and docking process. The style of the video is realistic, with a focus on the ship and its surroundings. The video does not contain any people or animals, and the focus is solely on the ship and its immediate environment.

Figure 6:

A vibrant magpie, its feathers a striking contrast of soft white and cool black, perches gracefully atop a slender branch of an ancient flower tree. The bird's delicate talons grip the bark with precision, as soft morning light filters through the leafless canopy, casting gentle shadows that dance around it. The magpie's head tilts slightly, its keen eye surveying the tranquil winter scene, while the rest of the forest lies quiet and still in the crisp, cool air of an early dawn.

Figure 7:

Extreme close-up of chicken and green pepper kebabs grilling on a barbeque with flames. Shallow focus and light smoke. vivid colours

Figure 8:

The video captures a wooden boat with coastal landscape. A small wooden boat is sailing on the sea surface. The wooden boat is composed of wooden boards with rich textures, with a rustic texture.

\textbf{Supplemental Material}

Figure 1:

A very cute little wild dog with a mouth and shiny fur, shining brightly towards the center of the picture, very lively. The dog takes a walk on the street with a natural flow of movements.

The video features a woman with long blonde hair, wearing a black and white dress, and large earrings. She is seated in a room with a white wall, which is adorned with various photographs. The woman is looking directly at the camera, and her expression is neutral. The room has a bright and airy feel to it. The style of the video is a combination of a personal interview and a motivational message.

The video features a red Tesla Model S driving on a highway with a mountainous landscape in the background. The car is sleek and modern, with a streamlined design and a large front grille. The highway is wide and well-maintained, with clear lane markings. The mountains rise majestically in the distance, their peaks shrouded in mist. The sky is a clear blue, and the sun is setting, casting a warm glow over the scene. The car is moving at a steady pace, and the driver appears to be focused on the road ahead. The overall style of the video is dynamic and energetic, capturing the thrill of driving a high-performance electric vehicle in a beautiful natural setting.

The video captures a serene coastal scene, showcasing the natural beauty of the area. The first frame shows a rocky shoreline with the calm ocean gently lapping against the rocks. The second frame reveals a small town nestled on the hillside, with houses and buildings scattered across the landscape. The third frame offers a closer view of the shoreline, with the waves crashing against the rocks, creating a dynamic contrast between the stillness of the ocean and the power of the waves. The overall style of the video is a blend of nature and human habitation, with a focus on the interplay between the land and the sea.

The video captures a sailboat journey across a vast body of water. The sailboat, with its white sail billowing in the wind, is the main focus of the video. The boat is seen from a high angle, providing a bird's eye view of its journey. The water around the boat is choppy, indicating a windy day. The sky above is a clear blue, suggesting a sunny day. The boat is moving towards the right side of the frame, indicating its direction of travel. The overall style of the video is a dynamic and adventurous depiction of a sailboat journey on a windy day.

Figure 2:

A fluffy, grey tabby kitten with bright, curious eyes and a white chest is playfully rolling on a patch of soft, green grass. The sun casts a gentle glow over its fur, emphasizing the subtle stripes and the warmth of its whiskers. The kitten pauses, sits up, and with a twitch of its little pink nose, it tilts its head inquisitively, capturing the essence of feline charm in a serene, outdoor setting.

Figure 3:

In a breathtaking close-up, a majestic gray wolf with a coat of realistic, silver-tipped fur stands regally against the lush, green tapestry of an ancient forest. Its eyes, deep and expressive, seem to hold the secrets of the wild, locking gazes with the viewer through the lens. With a mouth agape, the wolf appears to be caught in a silent howl or a gentle pant, breathing life into the serene scene. The atmosphere is one of unspoken grandeur, as the video encapsulates the raw, wild essence of the wolf amidst the tranquility of its untouched habitat.

The video captures the majestic Mont Saint-Michel, a historic abbey and fortification located on a small island in Normandy, France. The first frame shows the abbey from a distance, its grandeur accentuated by the clear blue sky. The second frame offers a closer view, revealing the intricate details of the abbey's architecture. The third frame provides a ground-level perspective, allowing viewers to appreciate the scale of the abbey and its surroundings. The video is a testament to the abbey's historical significance and architectural beauty, set against the backdrop of a serene day.

At the bottom of the screen is a green lake surface, rippling with the wind. A small wooden boat moved from the left side of the lake to the right side. In the middle are continuous mountains, towering and magnificent, not satisfied with the green forest, with snow mixed at the top of the mountains. The background is a blue sky.

Figure 4:

A bustling scene at the Piazza del Duomo in Milan, Italy. The focal point of the scene is the large archway entrance to the Galleria Vittorio Emanuele II, a historic shopping mall. The archway, constructed of white stone, stands majestically against the backdrop of a clear blue sky.   Flanking the entrance are two buildings, one on each side. These buildings, also made of white stone, are adorned with green awnings, adding a touch of color to the otherwise monochrome structure.   The piazza itself is teeming with life. People can be seen walking around, adding a dynamic element to the scene. In the background, a few trees can be spotted, providing a touch of nature amidst the urban setting.   Despite the scene being taken from a distance, the details of the archway and the surrounding buildings are clearly visible, showcasing the architectural grandeur of the landmark. The scene does not contain any discernible text. The relative positions of the objects suggest a well-planned architectural design, with the archway centrally located and the buildings symmetrically placed on either side. 

Figure 5:

The video captures a brown owl in flight, soaring through a forested area. The owl's wings are spread wide, showcasing its impressive wingspan. The owl's eyes are wide open, alert and focused on its surroundings. The background is a blur of green and brown, indicating the presence of trees and foliage. The owl's flight path is smooth and graceful, demonstrating the bird's agility and control. The overall style of the video is a close-up, slow-motion shot, allowing viewers to appreciate the details of the owl's flight and the beauty of its natural habitat.

Figure 6:

The video shows a chocolate cake being decorated with candy. The cake is placed on a white plate, and the candies are scattered on top of the cake. The candies are orange and yellow, and they are placed on top of the chocolate frosting. The cake is then drizzled with chocolate sauce, which is poured over the top of the cake. The sauce is thick and glossy, and it drips down the sides of the cake. The cake is then placed on a table, and the camera zooms in on the cake, showing the details of the decoration. The video is a close-up shot, focusing on the cake and the decoration. The style of the video is a food video, showcasing the process of decorating a cake.

A gleaming yellow cake, adorned with white frosting and delicate piping, sits at the center of a marble countertop. Above it, a cascade of vibrant pink strawberry juice, shimmering under the kitchen lights, gently pours from a height, drizzling down the sides of the cake, pooling slightly around the base, and creating a mesmerizing contrast between the yellow cake and the rosy liquid.

Figure 7:

The video shows a man with a beard standing next to a black truck. He is wearing a black T-shirt. In the first frame, he is pointing at the truck. In the second frame, he is opening the door of the truck. In the third frame, he is sitting inside the truck. The truck is parked in a lot with trees in the background. The man appears to be in the process of getting into the truck. The style of the video is casual and informal.

The video captures the majestic Neuschwanstein Castle, a 19th-century Romanesque Revival palace, perched on a rugged hill above the village of Hohenschwangau near Füssen in southwest Bavaria, Germany. The castle, a symbol of fairy tales and fantasy, is seen in three different frames, each showcasing its grandeur against the backdrop of the snow-capped mountains and the surrounding forest. The first frame offers a distant view of the castle, its multiple towers and turrets reaching towards the sky. The second frame provides a closer perspective, revealing the intricate details of the castle's architecture. The third frame offers a panoramic view of the castle, its towers and turrets standing tall against the backdrop of the snow-capped mountains and the surrounding forest. The video is a testament to the castle's historical significance and architectural beauty.

A close-up shot captures the innocent, yet adventurous expression of a young, blonde girl, her eyes a shimmering shade of blue. She's adorned in a classic khaki canvas top hat, casting a gentle shadow over her bright, curious eyes. The scene is one of quiet wonder, with the girl's face conveying a mix of mischief and wonder, as if she's about to embark on a grand, unknown journey.

An ancient beige white temple, crafted from marble with a subtle yellow patina, stands majestically amidst a serene landscape. It boasts an array of towering pillars, each meticulously carved, supporting a grand, beige white roof. The temple's weathered surface reveals the whispers of time, marked by intricate cracks and the subtle discolorations that speak to its storied past.

A small, light white fox with a fluffy tail and alert ears sits gracefully atop a lush green meadow, its head initially facing the viewer with a curious gaze. Gradually, the little fox tilts its head to the left, its eyes glinting with a mix of curiosity and wariness, as it surveys the surroundings. The soft sunlight filters through the nearby trees, casting dappled shadows that dance across the grass and the fox's fur, creating a tranquil and picturesque scene. The grassland in the background presents a bright yellow green color.

Figure 8:

A vibrant cluster of wildflowers, a kaleidoscope of reds, yellows, and purples, captures the foreground, their delicate petals swaying gently in the breeze. Beyond, a majestic range of snow-capped mountains stretches across the horizon, their peaks glistening under the midday sun, creating a breathtaking contrast against the clear blue sky. The flowers, a testament to the valley's vibrant life, and the distant mountains, symbols of enduring strength, together paint a picture of serene natural beauty.

A majestic white chrysanthemum, its circular petals meticulously wrapped around a vibrant yellow core, sways gracefully in a gentle breeze. Each petal, delicate and perfectly formed, shimmers with the morning dew, catching the light as the flower dances with the rhythm of the wind. The scene captures the delicate balance of nature, the floral beauty moving in harmony with the unseen forces around it, a serene moment of natural elegance.

A vibrant yellow hummingbird, its iridescent feathers shimmering in the soft sunlight, hovers with exquisite precision above a pristine white flower. The scene is captured in side profile, showcasing the bird's slender beak as it extends towards the flower's nectar-rich center, wings blurred by the rapidity of their motion, a mesmerizing dance of nature's delicate balance.

Figure 9: 

A fluffy Pomeranian dog weara light colored top hat made of cloth adorned with a pink satin bow tied neatly around the base of the hat. The dog has a light brown coat with white markings on its face and chest. The top hat, slightly tilted, gives the Pomeranian a charming and playful appearance, while the pink bow adds a touch of elegance and whimsy. The dog’s fur is well-groomed, showcasing its signature soft, voluminous coat that frames its small face.

Figure 10: 
A mouthwatering round cake, adorned with  a dusting of powdered sugar, and a generous drizzle of glossy, dark chocolate ganache. The cake itself is a masterpiece of delicate sponge layers, each separated by a rich, velvety buttercream frosting that peeks through the sides, inviting everyone to take a slice. The scene is set with a soft focus on the cake, positioned on a marble countertop, with gentle lighting casting subtle shadows to accentuate its delectable texture.

Figure 11:

The video captures a brown owl in flight, soaring through a forested area. The owl's wings are spread wide, showcasing its impressive wingspan. The owl's eyes are wide open, alert and focused on its surroundings. The background is a blur of green and brown, indicating the presence of trees and foliage. The owl's flight path is smooth and graceful, demonstrating the bird's agility and control. The overall style of the video is a close-up, slow-motion shot, allowing viewers to appreciate the details of the owl's flight and the beauty of its natural habitat.

The video shows a close-up of three chocolate-covered desserts with various toppings, including nuts and a red filling, on a white plate. The desserts are presented in a way that suggests they are being eaten or sampled, with one dessert partially cut into and a spoon nearby. The style of the video is simple and straightforward, focusing on the desserts without any additional context or background. The lighting is bright, highlighting the textures and colors of the desserts. The video is likely intended to showcase the desserts' presentation and appeal to viewers' appetites.

Figure 12:

The video features a large, gray teddy bear with a white nose and a blue ribbon around its neck. The bear is holding a small cake with a single candle on it. The bear is sitting in a room with a white wall in the background. The bear appears to be looking at the cake. The video is likely a celebration of a birthday or a special occasion. The bear's fur is soft and fluffy, and the blue ribbon adds a touch of color to the gray bear. The cake is small and white, with a single candle on top. The bear's eyes are closed, and it seems to be enjoying the moment. The room is brightly lit, and the white wall in the background provides a clean and simple backdrop for the scene. The bear is the main focus of the video, and its size and position in the frame make it the most prominent object. The cake is smaller in comparison, but it still stands out due to its bright color and the candle on top. The bear's position relative to the cake suggests that it is about to blow out the candle. The overall style of the video is simple and straightforward, with a focus on the bear and the cake. The lighting is bright and even, and the colors are soft and muted. The video does not contain any text or additional objects, and the focus is solely on the bear and the cake. 

The video shows a serene waterfront scene with two large, multi-story houses situated on a grassy hillside. The houses are constructed with a combination of stone and wood, featuring traditional architectural elements such as pitched roofs and bay windows. The houses are surrounded by lush greenery, including a variety of trees and shrubs, which add to the natural beauty of the setting.  In the foreground, there is a wooden dock extending into the water, with a small boat moored at the end. The boat appears to be a leisure vessel, possibly used for fishing or recreational purposes. The calm water reflects the houses and the surrounding landscape, creating a peaceful and idyllic atmosphere.  The style of the video is realistic and naturalistic, capturing the tranquility of the waterfront location with a focus on the architecture and the natural environment. The camera angles and movements are steady and unhurried, allowing the viewer to appreciate the details of the houses and the surrounding landscape. The lighting is soft and diffused, suggesting either an overcast day or a time when the sun is not directly shining on the scene. Overall, the video presents a picturesque and serene waterfront setting, likely intended to evoke a sense of relaxation and escape from the hustle and bustle of everyday life.

Figure 13:

Two pristine white flowers gently sway in the gentle breeze, their delicate petals catching the soft rays of sunlight that filter through the surrounding foliage. Each flower, with its subtle fragrance and intricate design, appears almost weightless as it dances gracefully in the air, creating a serene and tranquil atmosphere. The movement of these blossoms against the backdrop of lush greenery adds a dynamic element to the otherwise still scene, evoking a sense of peace and beauty in nature.

A fluffy Pomeranian dog weara light colored top hat made of cloth adorned with a pink satin bow tied neatly around the base of the hat. The dog has a light brown coat with white markings on its face and chest. The top hat, slightly tilted, gives the Pomeranian a charming and playful appearance, while the pink bow adds a touch of elegance and whimsy. The dog’s fur is well-groomed, showcasing its signature soft, voluminous coat that frames its small face.

Figure 14:

The video features a bald man with a beard and glasses, wearing a white shirt. He is standing against a blue background. The man appears to be speaking or presenting, as he is gesturing with his hands. The style of the video is professional and polished, suggesting it could be a corporate or educational video. The man's attire and the background give the impression of a formal or professional setting.

In a whimsical transformation, a sleek black cat with piercing green eyes sits gracefully on a windowsill, bathed in the warm glow of the afternoon sun. As the scene progresses, the cat blinks slowly, and with a shimmering aura, begins to morph seamlessly into a majestic bird, its fur transitioning into feathers of iridescent brown and white. The newly formed bird spreads its powerful wings, perched confidently on the same sill, now surveying the open sky with an air of freedom and wild spirit.

{
    \small
    \bibliographystyle{ieeenat_fullname}
    \bibliography{supple}
}